\documentclass[conference]{IEEEtran}
\usepackage{graphicx}
\usepackage{balance}  
\usepackage{caption}
\usepackage{color, colortbl}
\usepackage{comment}
\usepackage{booktabs}
\usepackage{enumerate}
\usepackage{tabularx}
\usepackage{bm}
\usepackage{multirow}
\usepackage{soul}
\usepackage{algpseudocode}
\usepackage{algorithm,algorithmicx}
\usepackage{subfigure}
\usepackage{amsfonts}
\usepackage{color}

\usepackage{amsthm}
\usepackage{url}
\usepackage{amsmath}
\usepackage{tikz}
\newcommand{\ballnumber}[1]{\tikz[baseline=(myanchor.base)] \node[circle,fill=.,inner sep=1pt] (myanchor) {\color{-.}\bfseries\footnotesize #1};}

\newcommand{\hang}[1]{\textcolor{red}{[\small FIXME: ~#1~]}}

\newtheoremstyle{mystyle}
{.1in}
{.1in}
{\itshape}
{}
{\bfseries}
{.}
{ }
{}%

\theoremstyle{mystyle}

\newcommand{\lda}{\textsc{ezLDA}}

\algnewcommand{\LineComment}[1]{\State \(\triangleright\) #1}
\algnewcommand\algorithmicswitch{\textbf{switch}}
\algnewcommand\algorithmiccase{\textbf{case}}
\algnewcommand\algorithmicassert{\texttt{assert}}
\algnewcommand\Assert[1]{\State \algorithmicassert(#1)}%

\algdef{SE}[SWITCH]{Switch}{EndSwitch}[1]{\algorithmicswitch\ #1\ \algorithmicdo}{\algorithmicend\ \algorithmicswitch}%
\algdef{SE}[CASE]{Case}{EndCase}[1]{\algorithmiccase\ #1}{\algorithmicend\ \algorithmiccase}%
\algtext*{EndSwitch}%
\algtext*{EndCase}%

\begin{document}



\title{{\lda}: Efficient and Scalable LDA on GPUs}




\author{\IEEEauthorblockN{Shilong Wang}
\IEEEauthorblockA{Department of Electrical and\\Computer Engineering\\
University of Massachusetts Lowell\\
Lowell, MA, USA 01854\\
Email: shilong\_wang@student.uml.edu
}
\and
\IEEEauthorblockN{Hang Liu and Anil Gaihre}
\IEEEauthorblockA{Department of Electrical and\\Computer Engineering\\
Stevens Institute of Technology\\
Hoboken, NJ, USA 07030\\
Email: hliu77@stevens.edu,\\agaihre@stevens.edu}


\and
\IEEEauthorblockN{Hengyong Yu}
\IEEEauthorblockA{Department of Electrical and\\Computer Engineering\\
University of Massachusetts Lowell\\
Lowell, MA, USA 01854\\
Email: hengyong\_yu@uml.edu}}

\maketitle

\begin{abstract}

Latent Dirichlet Allocation (LDA) is a statistical approach for topic modeling with a wide range of applications.
Attracted by the exceptional computing and memory throughput capabilities, this work introduces {\lda} which achieves efficient and scalable LDA training on GPUs with the following three contributions: 
First, {\lda} introduces three-branch sampling method which takes advantage of the convergence heterogeneity of various tokens to reduce the redundant sampling task. 
Second, to enable sparsity-aware format for both \textbf{D} and \textbf{W} on GPUs with fast sampling and updating, we introduce hybrid format for \textbf{W} along with corresponding token partition to T and {inverted index designs}. 
Third, we design a hierarchical workload balancing solution to address the extremely skewed workload imbalance problem on GPU and scale {\lda} across multiple GPUs. 
Taken together, {\lda} achieves superior performance over the state-of-the-art attempts with lower memory consumption. 

	
\end{abstract}

\section{Introduction}

Topic modeling is a type of statistical approach that reveals the \textit{latent} (i.e., unobserved) topics for a collection of documents (also referred to as corpus). LDA~\cite{LDAOriginal}, which \textit{carefully} chooses the Dirichlet distribution as the statistical model to formulate topic distributions, is one of the most popular topic modeling approaches that find applications in not only text analysis~\cite{textAnalysis,zhang2018zwift}, but also computer vision~\cite{computerVision}, recommendation system~\cite{recommendationSystem1, recommendationSystem2} and network analysis~\cite{network} among many others. Thanks to the practical implications, contemporary search engines rely upon LDA to handle billions of news with 10K of topics and 1000K of words~\cite{LDAStar}.


Recently, we also observe interesting interactions between LDA and popular deep learning models.
First, 
Functional and Contextual attention-based Long Short-Term Memory (FC-LSTM)~\cite{FC-LSTM} uses LDA to preprocess the data and feeds the corresponding results into LSTM model to improve the accuracy in a recommendation system. LDA can also cooperate with Convolutional Neural Network (CNN) model as a preprocessing method to deal with automobile insurance fraud problems~\cite{insuranceFraud}.
Second, logistic LDA~\cite{LogisticLDA}, which is a modified supervised LDA model, can achieve comparable accuracy with Syntax Aware LSTM (SA-LSTM)~\cite{SA-LSTM} on document classification with much shorter training time than SA-LSTM. BPLDA~\cite{BPLDA} can achieve comparable accuracy on classification and regression as deep learning with much shorter training time. {Compared with recent deep learning based natural language processing (NLP) tools, e.g.,} Embeddings from Language Models (ELMo)~\cite{ELMo} and Bidirectional Encoder Representations from Transformers (BERT)~\cite{BERT,QBERT}, LDA also presents a solid theoretical foundation which is absent for deep learning models.





As the size of NLP problems continues to rise, it becomes imperative for us to scale the training of LDA towards more computing resources, as well as accommodating larger corpus with more topics. 
{Graphics Processing Units (GPUs) exhibits remarkable performance over traditional CPU system and are hence widely applied on compute-intensive problems such as} deep learning~\cite{GPUChair1,GPUChair2,GPUSpeech,GPUDistributed,GPULargeScale} and graph~\cite{GPUGraph}. Towards expediting LDA training, GPUs are a tempting platform for two, if not more, reasons. First, modern GPUs yield extraordinary computing and data delivering capabilities, both of which are crucial for LDA training. Second, GPUs possess a thriving community with steady updates in both hardware and software support, which helps extend the impacts of LDA. 


Generally speaking, LDA encompasses three tensors and two tasks. First, the three tensors are: the token list T - an array of $<$wordId, docId, topicId$>$ triplets, a document-topic matrix (i.e., \textbf{D}) and a word-topic matrix (i.e., \textbf{W}). Second, the two tasks are sampling and updating. During sampling, LDA takes as input each specific token, i.e., \textit{t}, and relies on \textbf{D} and \textbf{W} to generate a new topic for this token. The intuition of sampling is that \textit{the probability of assigning new topic $t$ is positively correlated to the number of tokens for each topic of the document and word this $t$ belongs to}. During updating stage, {T}, \textbf{D} and \textbf{W} are updated to reflect the new topic generated for each token $t$.


\subsection{Related Work}

As one of the most popular topics in machine learning, LDA~\cite{LDAOriginal} has received enormous attentions. This section restricts our discussions to the projects that are closely related to {\lda}, that is, efficiency, scalability and GPU-based LDA. 


There mainly exist four directions to make LDA more efficient than the original attempt~\cite{LDAOriginal}, that is, sparsity-aware, Metropolis-Hasting (MH) and Expectation Maximization (EM) approaches, as well as the hybrid of them. \textbf{i)}. Sparsity-aware method utilizes the sparsity of word-topic and document-topic matrix to make the sampling time sublinear to number of topics K (detailed in Section~\ref{LDA Model}). SparseLDA~\cite{sparseLDA} pioneers this attempt. \textbf{ii)}. MH~\cite{MH} method generates a complex distribution by constructing a Markov Chain (MC) with a simple easy-to-sample distribution and update the topic with some acceptance rates at each step.
Since MH requires frequent random memory address to word-topic and doc-topic matrices, thus is not friendly for sparse matrix.
\textbf{iii)}. LDA*~\cite{LDAStar} uses sparsity-aware and MH samplers to deal with short and long documents separately.
The follow-up variations are~\cite{AliasLDA,LightLDA,warpLDA,SaberLDA,culdacgs,BIDMach,F+LDA}.
\textbf{iiii)}. EM~\cite{EPSGLD} divides the LDA training into E-step and M-step while the former is responsible for sampling and the latter for updating. 
Comparing with standard LDA sampling, EM can enjoy better parallelism because frequent random memory access to update word-topic and document-topic matrices during sampling can be avoided.
Large-scale training is another important field for LDA considering real-world corpus often contains billions of tokens. 
LightLDA~\cite{LightLDA} leads this effort. Particularly, it trains LDA model with 1 million topics and 1 million words on eight machines via data parallelism (corpus partition) and model parallelism (word-topic matrix splitting).
While LightLDA allows both \textbf{D} and \textbf{W} to be sparse, it relies upon hash table for fast sampling, which is a hardership for GPU because collision handling in hash table remains elusive on GPUs~\cite{ashkiani2018dynamic}.
This concern is evident by SaberLDA~\cite{SaberLDA} which only stores \textbf{D} in sparse format for fast sampling. 
cuLDA~\cite{culdacgs} further attempts this challenge on multi-GPU but ends up with identical strategy as SaberLDA except scaling the sampling towards multiple GPUs. 
As we will evaluate in Table~\ref{table:mem}, only allowing \textbf{D} to be sparse will greatly hinder the scalability of LDA.
Last, for GPU-based LDA, which is the interest of this work, we find very few efforts. \textit{Yan et al}~\cite{CGSGPU} implements collapsed Gibbs sampling~\cite{CGSOriginal} and collapsed variational Bayesian~\cite{collapsedVB} on GPU. BIDMach~\cite{BIDMach} toolkit implements Monte Carlo Expectation Maximization (MCEM)~\cite{MCEM} method on GPU without much GPU specific optimizations thus ends up with moderate performance. SaberLDA~\cite{SaberLDA} proposes the PDOW (partition by document, order by word) strategy to reduce random memory access. Warp-based sampling is also adopted, which means using a warp to process a token and a block to process a word, to avoid thread-divergence and uncoalesced memory access. Further, cuLDA~\cite{culdacgs} scales LDA to multiple GPUs based on collapsed Gibbs sampling with similar optimizations as SaberLDA. In summary, \textbf{the curse of GPU-based LDA is the limited number of topics because they have to store \textbf{W} in dense format} - larger topics will exhaust the limited memory space of GPUs. 



\vspace{-.05in}
\subsection{Contributions}

This paper introduces {\lda}\footnote{{\lda}, pronounced as ``easy LDA'', implies that this project achieves efficiency and scalability without the involvement of users.}, an efficient and scalable LDA project that trains LDA across multiple GPUs. Notably, {\lda} can train LDA on UMBC dataset within 700 seconds while supporting the unprecedented 32,768 topics on merely one V100 GPU~\cite{nvidia_V100}. This achievement is not possible without the following contributions: 

First, {\lda} introduces one more direction to make LDA efficient, i.e., the three-branch sampling method which takes advantage of the convergence heterogeneity of various tokens to reduce the redundant sampling task. 
While the convergence heterogeneity is promising, the caveat is that one cannot simply avoid sampling a token because its topic remains unchanged for a number of iterations, as detailed in Figure~\ref{fig:remerge}.
Inspired by our key observation that the majority of the tokens often fall in the top popular topics, we single out these popular topics as the third sampling branch in addition to the traditional two branches (detailed in Section~\ref{LDA Model}). During sampling, we introduce an algorithm to accurately estimate whether this token will remain in the top popular topics thus be skipped or not. {Meanwhile, in order to minimize the overhead of three-branch sampling, we introduce processing by both word and document strategy along with inverted index, and top topics pair-storage.} Our evaluation shows three-branch sampling can avoid sampling over 70\% of the tokens {with negligible overhead after 100 iterations} on large datasets, PubMed and UMBC.

Second, to enable sparsity-aware format for both \textbf{D} and \textbf{W} on GPUs with fast sampling and updating upon this format, we introduce hybrid format for \textbf{W} along with corresponding token partition to T and inverted index designs.
Particularly, we store \textbf{W} in sparse and dense hybrid format and \textbf{D} in sparse format. During sampling, we will keep a canonical copy of the dense format of \textbf{W}, which accounts for the majority of the tokens but with very few number of words, in GPU memory to cache the updates. 
After sampling, we will update both the sparse part of \textbf{W} and \textbf{D}. Since sparse part of \textbf{W} holds very few tokens, the update is trivial. Pair-storing the row index and value of \textbf{D} is also adopted for fast sampling. For rapid update of \textbf{D}, we, again, leverage the inverted index of \textbf{D} to navigate through the token list T for tokens of interest. 
We also notice that SaberLDA~\cite{SaberLDA} has attempted sparsity aware LDA but ends up with only sparse \textbf{D} which cannot solve memory exhaustion problem caused by dense \textbf{W} when vocabulary and topic size become too large. Consequently, {\lda}, as shown in Table~\ref{table:mem}, doubles the space saving over SaberLDA thus supports models that SaberLDA cannot.


Third, we improve the scalability of {\lda} across GPU threads and GPUs. For single GPU, we introduce hierarchical workload management to ultimately balance the workload which consists of two optimizations. Specifically, we first use atomic operation to dynamically decide which word should be assigned to which GPU warp thanks to light-weighted GPU atomic operation~\cite{gaihre2019xbfs}. 
Further, we propose to split the extremely large words for better workload balancing. To efficiently combine dynamic inter-word scheduling and large word splitting design, we introduce efficient indexing to achieve light-weighted workload management.
Towards multi-GPU support, we propose to cache \textbf{W}, partitioned \textbf{D} and partitioned T in GPU memory to further boost the performance of {\lda}.

The novelty of this paper is that we introduce the \textit{efficient and scalable techniques} to achieve fast LDA training. Particularly, to the best of our knowledge, our three-branch sampling is the first successful design to exploit the convergence heterogeneity of various tokens for fast LDA sampling. We also shed lights on the possibility of using inverted index to achieve sparsity-aware LDA training where both \textbf{W} and \textbf{D} are sparse. It is also important to note that this paper strives to make sense of the complicated mathematical designs of LDA with an intuitive example which will also benefit the community.


\subsection{Organization}

The rest of the paper is organized as follows: Section~\ref{sec:background} explains the background. Section~\ref{sec:branch} presents the novel three-branch sampling design.
Section~\ref{sec:sparse} discusses sparsity-aware LDA and scalable LDA is introduced in Section~\ref{sec:workload}. 
Section~\ref{sec:eval} evaluates this work and we conclude in Section~\ref{sec:conclusion}.


\begin{figure}[t]
	\centering
	\includegraphics[width=0.4\textwidth]{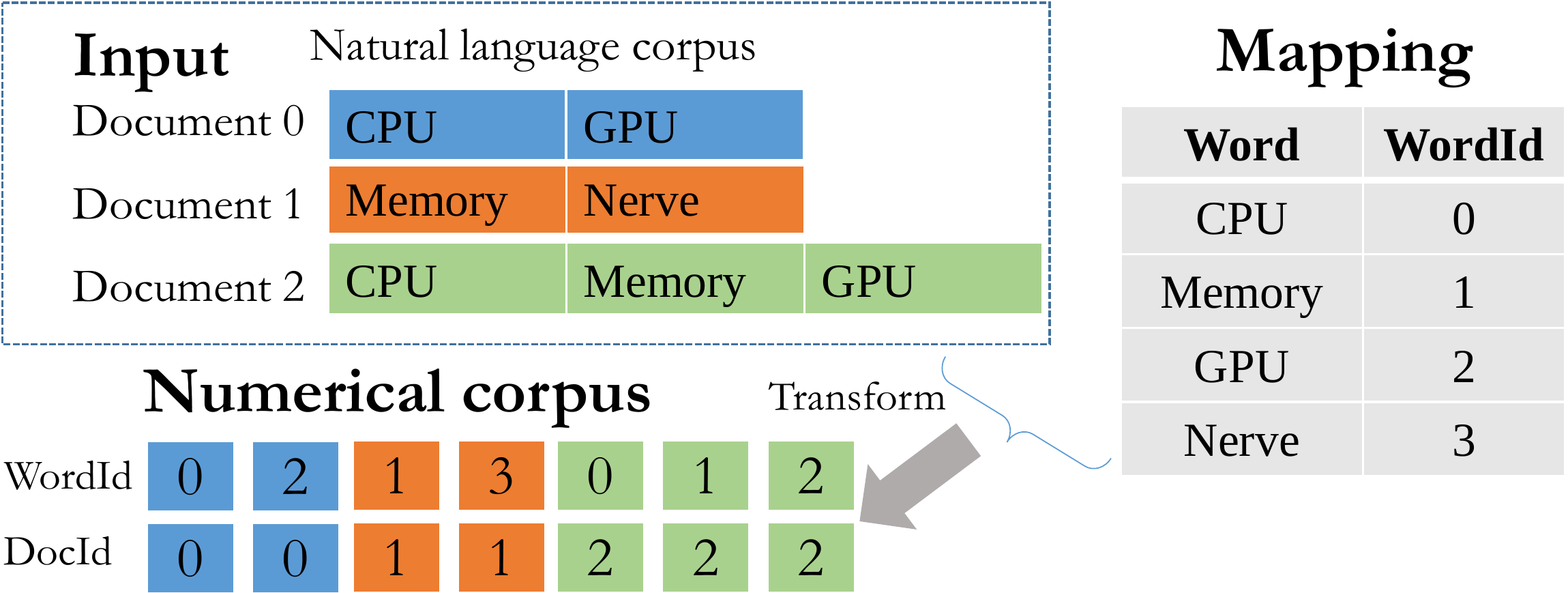}
	\caption{The running example used in this paper.
	}
	\vspace{-.3in}
	\label{fig:LDAExample}
\end{figure}

\section{Background and Challenges}
\label{sec:background}

\begin{figure*}[t]
	\centering
	\includegraphics[width=0.92\textwidth]{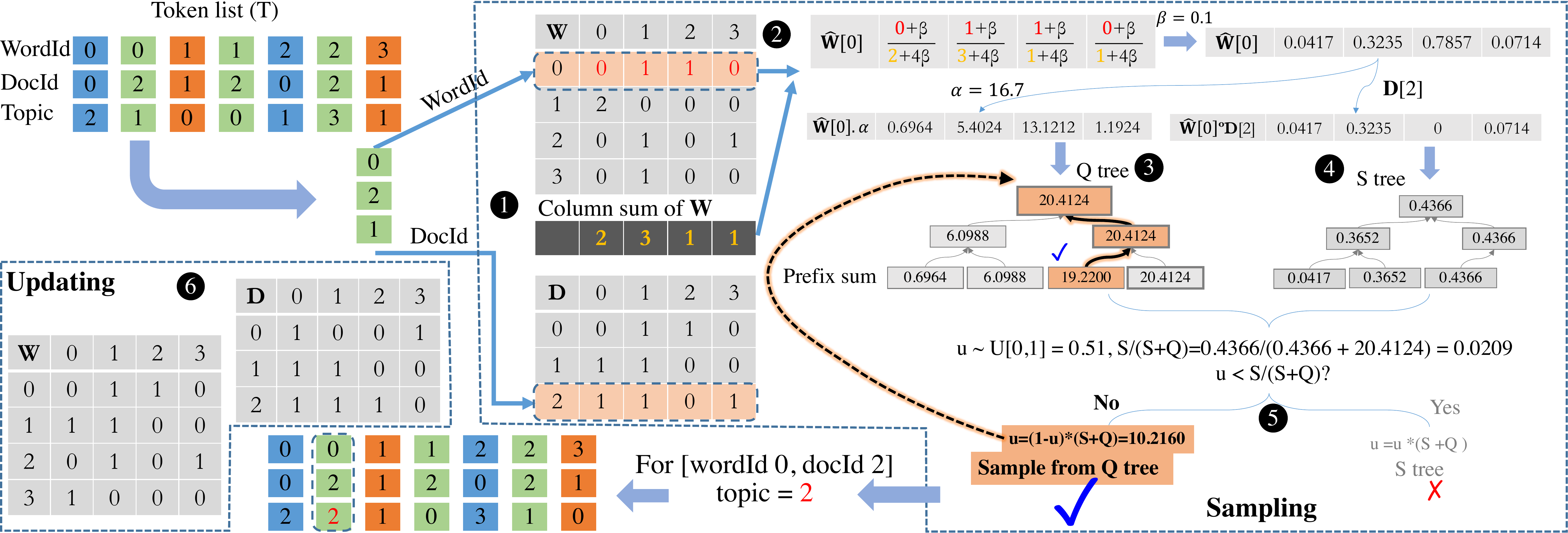}\vspace{-.1in}
	\caption{\protect{Two-branch sampling of one token for the corpus in Figure}~\ref{fig:LDAExample}. {Better viewed in color.}\vspace{-.2in} }
	\label{fig:LDASimple}
\end{figure*}

\subsection{General Purpose GPUs}
\label{GPU Basis}

Without loss of generality, this section uses Volta GPUs as an example to illustrate the essential backgrounds about modern GPUs, mainly from three aspects, that is, processor, memory and programming primitives. For more details about GPUs, we refer the readers to~\cite{GPUPaper1}. 

The streaming multiprocessor (SMX), which consists of several CUDA cores, is a basic processing chip for GPUs. For instance, Nvidia Tesla V100 GPUs~\cite{nvidia_V100} contain 80 SMXs, each of which has 64 single-precision CUDA cores and 32 double-precision units and a 96 KB shared memory/L1 cache and 65,536 registers.
V100 also features 6MB L2 cache and 16 GB global memory, which is shared by all SMXs. 
Similar to CPU, the memory access latency increases from register to shared memory, further to L2 cache and global memory. 

With massive CUDA cores, GPUs can run a large number of threads. Contemporary GPUs thus manage threads by warps, which is a group of 32 consecutive threads.
It is important to mention that a warp of threads is executed in Single Instruction Multiple Thread (SIMD) fashion, which is also one of the most representative features of modern GPUs.  
In terms of programming primitive, recent GPUs provide several warp-level primitives such as \textit{\_\_shfl()} and \textit{\_\_ballot()} for fast intra-warp communication.

%
%
%
%


\subsection{LDA Algorithm and Theory}
\label{LDA Model}





Before explaining LDA designs, Figure~\ref{fig:LDAExample} describes how to transform a real world natural language corpus into numerical corpus which can be used by LDA. 
Particularly, for a natural language corpus which consists of three documents with 2, 2 and 3 tokens, respectively, the preprocessing step will extract the unique words and assign each of them a specific wordId in mapping. 
This step is necessary because identical word might appear repeatedly, where each occurrence is called a token, e.g., memory appears in both documents 1 and 2.
After transformation, we arrive at the numerical corpus.



\textbf{Algorithm}.
LDA is a three-layer Bayesian model, that is, each document is viewed as a distribution of topics and each topic is further deemed as a distribution of vocabulary. 
For a given token, a new topic can be generated based on these two distributions. So, during training, two matrices are involved, i.e., \textbf{D} and \textbf{W}. While detailed theory behind why LDA would work can be found in~\cite{LDAOriginal}, this paper focuses on the algorithm.

For each token, ESCA~\cite{ESCA} - one of the popular LDA version -  assigns this token to topic $k$, i.e., $p(k)$ through the {following equation}:

\begin{equation}
\label{con:sampling prob}
{
p(k)\propto\underbrace{(\textbf{D}[d][k]+\alpha)}_{part 1}\cdot\frac{\textbf{W}[v][k]+\beta}{\underbrace{(\sum_{v=1}^{V}\textbf{W}[v][k]+V\beta)}_{part 2,\ \widehat{\textbf{W}}[v][k]}}},
\end{equation}
where $\alpha$ and $\beta$ are two constant hyper parameters. Similarly to~\cite{SaberLDA,culdacgs}, we adopt $\alpha=50/K$ and {$\beta=0.01$} for {\lda}, where $K$ is the total number of topics. $\textbf{D}[d][k]$ is the number of tokens in document $d$ that belongs to topic $k$ in \textbf{D}. Similarly, $\textbf{W}[v][k]$ is the number of tokens of word $v$ that belongs to topic $k$ in \textbf{W}. 


	

The intuition of Equation~(\ref{con:sampling prob}) is that, for token $t$ that belongs to word $v$ and document $d$, if more tokens from document $d$ and word $v$ fall in topic $k$, LDA will be more likely to assign topic $k$ to this token $t$, that is, $\textbf{D}[d][k]+\alpha$ and $\textbf{W}[d][k]+\beta$ will be larger. Further, the total number of tokens in $v$ - $\sum_{v=1}^{V}\textbf{W}[v][k]+V\beta$ - is negatively correlated.

Defining part 2 of Equation~(\ref{con:sampling prob}) as $ \widehat{\textbf{W}}[v][k]$, which can be regarded as the normalized version of $\textbf{W}$ matrix, we get:

%

\begin{equation}
{
p(k)\propto(\textbf{D}[d][k]+\alpha)\cdot\widehat{\textbf{W}}[v][k],
}
\label{con:W hat}
\end{equation}


It is important to note that we choose to extend ESCA~\cite{ESCA} because
 ESCA is sparsity-aware, which means the time complexity is sub-linear with respect to the number of topics.
ESCA achieves this sparsity-aware goal by decomposing the part 1 of Equation~(\ref{con:sampling prob}) into two separate terms. 
 So Equation~(\ref{con:sampling prob}) can be rewritten in the following format:

\begin{equation}
{
p(k)\propto =\underbrace{\textbf{D}[d][k]\cdot\widehat{\textbf{W}}[v][k]}_{p_{s}(k)}+\underbrace{\alpha\cdot\widehat{\textbf{W}}[v][k]}_{p_{q}(k)},
}
\label{con:optimized sampling prob}
\end{equation}
Equation~(\ref{con:optimized sampling prob}) can be further written into vector format:

\begin{equation}
\textbf{p}\propto(\textbf{D}[d] +\bm{\alpha} )\circ\widehat{\textbf{W}}[v]=\underbrace{\textbf{D}[d]\circ\widehat{\textbf{W}}[v]}_{\textbf{p}_{s}, \mathrm{\ or\ S\ tree}}+\underbrace{\bm{\alpha}\circ\widehat{\textbf{W}}[v]}_{\textbf{p}_{q},\mathrm{\ or\ Q\ tree}},
\label{con:optimized sampling prob vector}
\end{equation}
where $\bm{\circ}$ is the Hadamard Product (HP) operator. $\widehat{\textbf{W}}[v]$ is the normalized $v$-th row of $\textbf{W}$. $\textbf{D}[d]$ is $d$-th row of $\textbf{D}$. $\bm{\alpha}$ is a vector with all elements to be $\alpha$. 


Finally, ESCA defines S and Q as the sum of $\textbf{p}_s$ and $\textbf{p}_q$, respectively, 
sampling process of LDA 
 becomes as follow. Note, we term this sampling method as two-branch because it has S and Q two branches. 
 \begin{itemize}
     \item Generating a random number $u\sim U[0,1]$.
     \item Generating the new topic by
    $\begin{cases}
       \textbf{p}_s, & \text{if $u \le \frac{S}{S+Q}$;}\\
      \textbf{p}_q, & \text{otherwise.}
    \end{cases}$ 
 \end{itemize} 



\textbf{Example}. Figure~\ref{fig:LDASimple} presents one iteration of LDA on the same corpus as shown Figure~\ref{fig:LDAExample} with randomly assigned topics. During initialization (\ballnumber{1}), one will generate the \textbf{W} and \textbf{D} matrices from the token list T, where \textbf{W} and \textbf{D} are document-topic and word-topic matrices, respectively. Particularly, the dotted box in \textbf{W} means the document 0 has 0, 1, 1 and 0 tokens for topics 0, 1, 2 and 3, respectively. Similarly, the dotted box in \textbf{D} means that word 2 has 1, 1, 0 and 1 tokens for topics 0, 1, 2 and 3, respectively.
Note, the column sum of \textbf{W} is also computed, as shown below \textbf{W}, which means, in total, we have 2, 3, 1 and 1 tokens for topics 0, 1, 2 and 3, respectively. 

LDA training encompasses two steps, i.e., sampling and update. Further for sampling, LDA uses either Q or S tree to conduct sampling thus is called two-branch sampling. For the second token of the first word from the token list T - \{0, 2, 1\}, we follow Equation~(\ref{con:sampling prob}) to compute the $\widehat{\textbf{W}}[0]$ as \{$\frac{0+\beta}{2+4\beta}$, $\frac{1+\beta}{3+4\beta}$, $\frac{1+\beta}{1+4\beta}$, $\frac{0+\beta}{1+4\beta}$\} (\ballnumber{2}). Since $\alpha=50/3=16.7$ and $\beta=0.01$, we obtain $\widehat{\textbf{W}}[0]\circ\bm\alpha$ as \{0.0818, 5.5477, 16.2190, 0.1603\} and $\widehat{\textbf{W}}[0]\circ \textbf{D}[2]$ as \{0.0049, 0.3322, 0, 0.0096\}. Conducting prefix-sum~\cite{sengupta2007scan} of $\widehat{\textbf{W}}[0]\circ\bm\alpha$ and $\widehat{\textbf{W}}[0]\circ \textbf{D}[2]$, we arrive at the ranges of \{0.0818, 5.6295, 21.8485, 22.0088\} and \{0.0049, 0.3371, 0.3467, 0.3467\} for the Q and S tree, respectively.




The rule of tree construction is that the parent node should be the larger one of the two child nodes. Using the first two pairs of Q tree as an example, 5.6295 is the parent node of \{0.0818, 5.6295\} (\ballnumber{3}). Similarly for the rest of Q tree and S tree construction (\ballnumber{4}). The \ballnumber{5} step draws a random number from uniform distribution $U[0, 1]$, u = 0.51 in this case, and compares it against $\frac{S}{S+Q}$ to decide which tree to sample in order to derive a new topic for this token. Since 0.51 is not smaller than $\frac{S}{S+Q}$, we use Q tree to conduct the sampling by adjusting $u = (1 - u)\cdot(S+Q)$ = 10.9542 and descending the Q tree to arrive at new topic = 2. Following this way, LDA will update the topics for all the tokens T, subsequently the \textbf{D} and \textbf{W} matrices (\ballnumber{6}).

Since T is sorted by wordId, we only need to construct Q tree once for all the tokens of the same word. S tree construction, in contrast, needs to be done more frequently because adjacent tokens often come from different documents, as shown in Figure~\ref{fig:LDASimple}.


\textbf{Evaluation metric}. We use log-likelihood per token (LLPT), also known as negative logarithm of perplexity, as the parameter to evaluate the converge of LDA. 

\begin{align}
LLPT=\frac{1}{N}\sum_{n=1}^{N}(log_{2}\sum_{k=1}^{K}(\frac{\textbf{D}[d][k]+\alpha}{\sum_{k=1}^{K}\textbf{D}[d][k]+K\alpha}\cdot\\\nonumber
\frac{\textbf{W}[v][k]+\beta}{\sum_{v=1}^{V}\textbf{W}[v][k]+V\beta})),
\label{con:Perplexity}
\end{align}
where N is the total number of tokens in this corpus. The rule is that LLPT should increase and gradually become stable when computation proceeds iteratively.
\subsection{LDA Challenges}
\label{subsec:obsv}
\textbf{Challenge \#1. LDA sampling presents high time complexity}. For each token, one needs to conduct topic number of floating point multiplication and addition operations in order to get a new topic. Assuming we have N tokens and K topics, the time complexity of sampling in one iteration is as high as $\mathcal{O}(N\cdot K)$. 
As a result, our three-branch sampling (discussed in Section~\ref{sec:branch}) will largely reduce the computation complexity of sampling. 

\textbf{Challenge \#2. LDA presents high space complexity.}
As shown in Figure~\ref{fig:LDASimple}, LDA allocates three data structures: T, \textbf{D} and \textbf{W}, which consume $\mathcal{O}(N)$, $\mathcal{O}(M\cdot K)$ and $\mathcal{O}(V\cdot K)$ space, respectively, where M, V and K are number of documents, words and topics, respectively. As we will discuss in Section~\ref{sec:sparse}, {\lda} introduces sparsity-aware compression and computation to save space for both \textbf{D} and \textbf{W}.


\section{Three-Branch Fast Sampling}
\label{sec:branch}



This section discusses two important observations, our three-branch sampling and implementation optimizations that lower the overhead of three-branch sampling. 
\begin{figure}[h]
	\centering
	\includegraphics[width=0.4\textwidth]{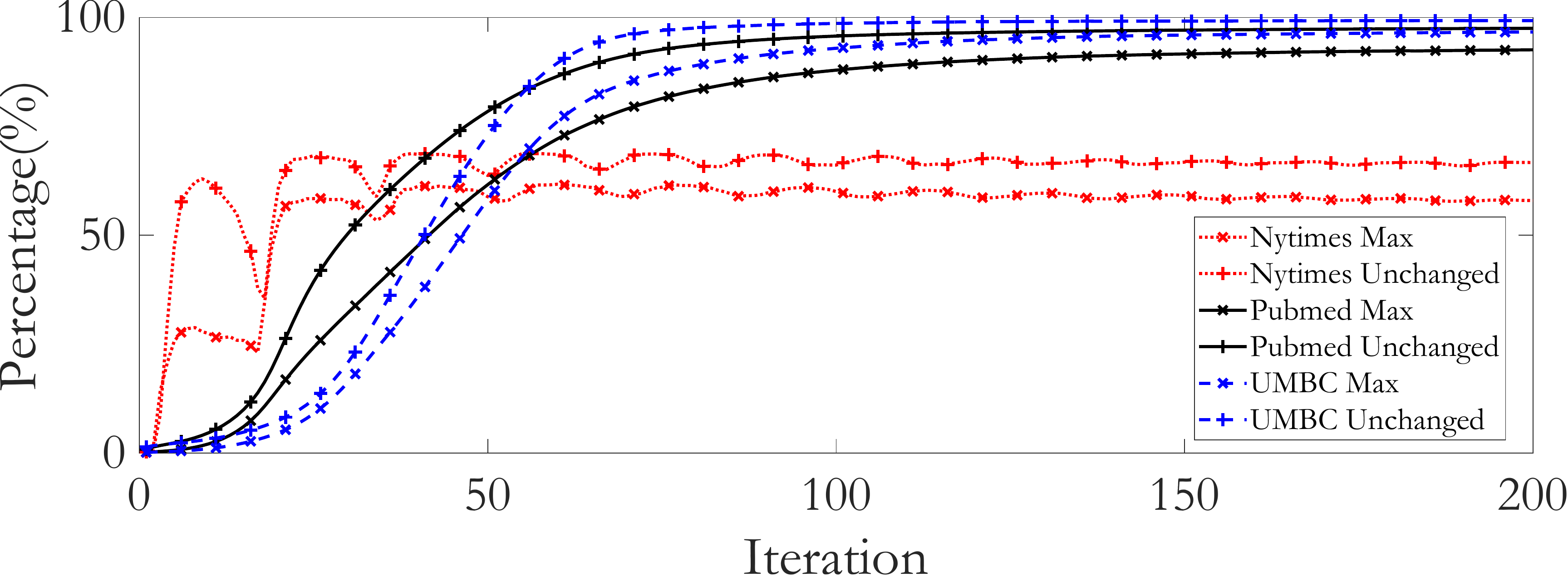}\vspace{-.1in}
	\caption{{Percentage of tokens with unchanged topic and tokens corresponding to the max topic. }
	\vspace{-.15in}}
	\label{fig:maxUnchanged}
\end{figure}

\begin{figure*}[t]
	\centering
	\includegraphics[width=\textwidth]{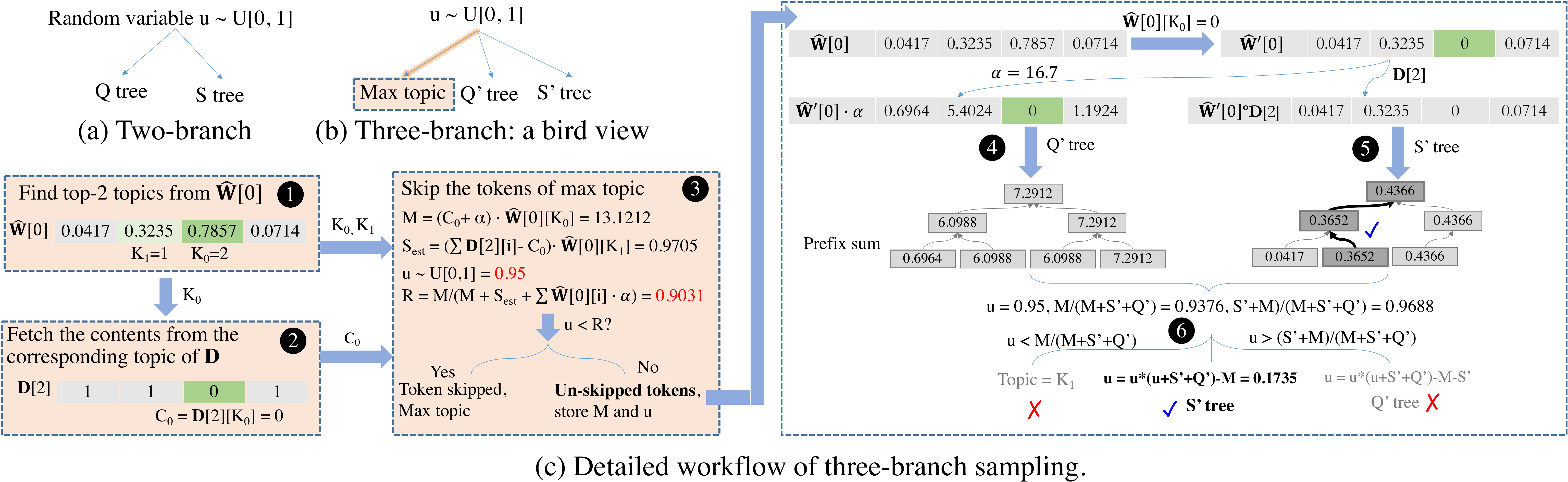}	
	\caption{{{\lda} three-branch sampling: (a) Two-branch vs (b) {\lda} three-branch sampling, a bird view and (c) Detailed workflow of three-branch sampling.} 
	\vspace{-.2in}
	}
	\label{fig:newMethod}
\end{figure*}

\subsection{Observations}
\label{subsec:obsv2}

This section makes the following two important observations that inspire our three-branch sampling. 

First, different tokens converge at dissimilar speeds. As shown in Figure~\ref{fig:maxUnchanged}, when iteration proceeds, more and more tokens experience unchanged topics. In other words, some tokens converge earlier and some later. For instance, at 50-th iteration, over 70\% of the tokens keep their topic unchanged in PubMed dataset.

Second, the majority of the tokens tend to converge to the most popular topic. This observation is self-explanatory - because a topic contains more tokens, and becomes the most popular topic. In fact, Figure~\ref{fig:maxUnchanged} quantitatively showcases this observation. In particular, more than 60\% of the tokens converge to the most popular topic in PubMed dataset at 50-th iteration.

The first observation implies that we can reduce the sampling workload for early converged tokens. However, since reducing the sampling task needs extra checking operations, this might incur significant overhead. Fortunately, our second observation further suggests that only focusing on the most popular topic would be adequate, which lowers the overhead.

\subsection{Three-Branch Sampling}




Since traditional two-branch sampling cannot leverage our observations in Section~\ref{subsec:obsv2} for workload reduction, we introduce three-branch sampling which singles out the most popular topic as one more branch. Below we discuss the theoretical soundness and implementation details of this design. Note, we cannot simply avoid sampling all the tokens from the most popular topic because, as discussed in Section~\ref{subs:dropping}, very few tokens from the most popular topic might change their topic, though more tokens will converge to the most popular topic.

\textbf{Theoretical soundness}. 
Our three-branch sampling rewrites Equation~(\ref{con:optimized sampling prob}) into the following format: 
\begin{equation}
\textbf{p}\propto\underbrace{\textbf{D}[d]\circ\widehat{\textbf{W}}'[v]}_{\textbf{p}_{s}}+\underbrace{\bm{\alpha}\circ\widehat{\textbf{W}}'[v]}_{\textbf{p}_{q}}+\underbrace{(\textbf{D}[d]+\bm{\alpha})\circ \widehat{\textbf{W}}[v]^m}_{\textbf{p}_{m}},
\label{con:fast converge}
\end{equation}
where $\widehat{\textbf{W}}'[v]$ is derived by setting the maximum entry of $\widehat{\textbf{W}}[v]$ as 0. On the contrary, $\widehat{\textbf{W}}[v]^{m}$ is achieved by setting all except the maximum entry of $\widehat{\textbf{W}}[v]$  as 0. 

Consequently, $\textbf{p}_{m}$ has only one non-zero entry which corresponds to the most popular topic. 
As shown in the left of Figure \ref{fig:newMethod}(a), traditional two-branch sampling approach conducts sampling from two branches -- either S or Q tree. Particularly, S and Q trees are constructed from $\textbf{p}_s$ and $\textbf{p}_q$ in Equation~(\ref{con:optimized sampling prob vector}), respectively. The proposed three-branch sampling, as shown in the right of Figure~\ref{fig:newMethod}(a), consists of three branches. That is, S' and Q' trees which are constructed from $\textbf{p}_s$ and $\textbf{p}_q$ in Equation~(\ref{con:fast converge}), respectively, and the max topic branch which is the $\textbf{p}_{m}$ in Equation~(\ref{con:fast converge}).





%

{\textbf{Three-branch sampling}}
is exemplified by Figure~\ref{fig:newMethod}(b). For each unique word, {\lda} first finds the top-2 topics in $\widehat{\textbf{W}}[v]$, which are $K_1=3$ and $K_2=1$ (\ballnumber{1}). Here ``top-2 topics" means these topics correspond to top-2 largest values, 4.5 and 0.3, in $\widehat{\textbf{W}}[v]$. Then given the third topic is the most popular one, we will extract the number of tokens from the same index in $\textbf{D}[d]$, that is, 2 (\ballnumber{2}). Consequently, $K_1=3, K_2=1$ and $C_1=2$. Afterwards, as shown in \ballnumber{3} of Figure~\ref{fig:newMethod}(b), we can calculate $M=16.22$ and $S_{est}=2.91$ and generate $u\sim U[0, 1]$. Compare $u$ against $\frac{M}{M+S_{est}+Q'}$ to determine whether this token remains in the most popular topic. If yes, this token will not involve in the following steps and corresponding topic will be updated to be $K_1$. Otherwise, store $M$ and u for this un-skipped token. Finally, we will execute the remaining steps (\ballnumber{4}, \ballnumber{5} and \ballnumber{6}), which are similar to two-branch sampling except following two differences: First, these steps only need to be done for the remaining un-skipped tokens. As training goes, more and more tokens are skipped and linear time decrease will be introduced. Second, max topic is singled out and considered separately. So $\widehat{\textbf{W}}[K_1]$ should be set to be zero and final sampling will include an additional M branch, as shown in the bottom right of Figure~\ref{fig:newMethod}(b), even after construction of S' tree, we can still avoid the sampling if $u<\frac{M}{M+S'+Q'}$. Besides, the Q' tree and S' tree (\ballnumber{4} and\ballnumber{5}) constructions are the same as two-branch sampling method.

\textbf{$S_{est}$ computation}. 
In order to skip as many tokens as possible, {\lda} needs to make $ S_{est} $ as close to S' as possible. {Meanwhile, to ensure theoretical soundness, we must also make sure tokens that go to 'Yes' branch in step} \ballnumber{3} must belong to the left branch in step \ballnumber{6}, which requires S' not greater than $ S_{est} $.
We use the following inequality to extract the $S_{est}$ and calculate M.
Assuming $\widehat{\textbf{W}}[v]$ is sorted in descending order:

\begin{equation}
\begin{aligned}
\widehat{\textbf{W}}[v] = [a_1, a_2, ..., a_n],\\
\textbf{D}[d] = [b_1, b_2, ..., b_n].
\end{aligned}
\label{con:}
\end{equation}
This means $a_{i}>a_{j}$ if $i>j$. Thus, maximum topic branch is: 
\begin{equation}
\begin{aligned}
M= a_{1}\cdot(b_{1}+\alpha).
\end{aligned}
\label{con:MValue}
\end{equation}
Given
\begin{flalign}
S'&= \widehat{\textbf{W}}[v]\cdot\textbf{D}[d] - M+a_{1}\cdot\alpha \\\nonumber
&=(a_{2}\cdot b_{2}+ a_{3}\cdot b_{3}+\cdots + a_{n}\cdot b_{n})\\\nonumber
&<\sum_{2\le i\le g}a_i\cdot b_i+a_{g+1}\cdot\sum_{g < i\le n}b_i,
\label{con:Srelation}
\end{flalign}
we hence propose:
\begin{equation}
\begin{aligned}
S_{est}=\sum_{2\le i\le g}a_i\cdot b_i+a_{g + 1}\sum_{g < i\le n}b_i.
\end{aligned}
\label{con:Sest}
\end{equation}
where $g \ge 1$ controls the \textbf{accuracy and cost} of the estimation. That is, the larger the value of $g$, the higher the cost, as well as the accuracy {between S' and $S_{est}$}.

\textbf{Parameter tuning}. 
First, {the choice of g} is a trade-off between benefit and overhead. {\lda} uses $g = 2$ because we can achieve significant better performance than g=1 with similar overhead after our optimization in Section \ref{sec:overhead}.
\subsection{Optimizations for Three-Branch Sampling}
\label{sec:overhead}

While three-branch sampling can avoid expensive S' tree constructions and sampling for all the skipped tokens, it also introduces three more steps, i.e., \ballnumber{1}, \ballnumber{2} and \ballnumber{3} as shown in Figure~\ref{fig:newMethod}(b). 

\begin{figure}[h]
	\centering
	\includegraphics[width=0.47\textwidth]{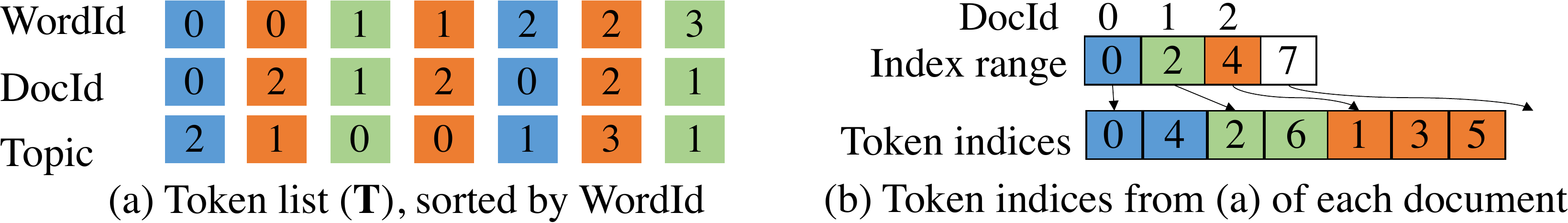}\vspace{-0.05in}
	\caption{Inverted index for document.\vspace{-.1in}}
	\label{fig:update}
\end{figure}

Across all the steps, the cost for steps~\ballnumber{1} and~\ballnumber{3} is negligible. For step~\ballnumber{3}, the workload is simple. For step~\ballnumber{1}, the reason lies in that the token list (i.e., T) is sorted by wordId, as shown in Figure~\ref{fig:update}(a), three-branch sampling only needs to find the top topics, that is, $K_1$ and $K_2$ pair in Figure~\ref{fig:newMethod}(b) -  once for all the tokens falling to the same word $v$. But also because T is sorted by wordId, step \ballnumber{2} would take significant amount of time if we want to find the corresponding $C_1$ and $C_2$ pair across all documents for each $v$ right after we find $K_1$ and $K_2$ pair.

In order to combat this overhead, {\lda} designs processing by word and document for $K_1$ and $K_2$ pair, and $C_1$ and $C_2$ pair, respectively. While processing by word is straightforward because T is sorted by word, processing by document turns out to be challenging. In this context, we introduce an inverted index for each document which stores the indices of tokens belonging to each document. This inverted index idea adopts Compressed Sparse Row (CSR) format~\cite{press2007numerical,chen2018locality} to store the indices of the tokens for each document. As shown in Figure~\ref{fig:update}(b), indices \{0, 4\}, \{2, 6\} and \{1, 3, 5\} are from documents 0, 1 and 2, respectively.

Scanning through the inverted index, we can fetch the corresponding row of \textbf{D}, as well as all the tokens of the same document easily. Note, in this processing by document design for $C_1$ and $C_2$ pair, we need to first write all the $K_1$ and $K_2$ pairs for all tokens into global memory and load them back for computation. However, this cost is way lower than we conduct processing by word for both $K_1$ and $K_2$ pair and $C_1$ and $C_2$ pair. Particularly, processing by word for $C_1$ and $C_2$ pair needs to repeatedly scan through the token list and search for $C_1$ and $C_2$ for each token because tokens of the same document are not stored together. It is also worthy to note that we combine $K_1$ and $K_2$, $C_1$ and $C_2$ pairs into a single value $K_{12}$ and $C_{12}$, where the higher half bits of them store $K_1$ and $C_1$, and the lower half bits store $K_2$ and $C_2$ respectively, in order to further reduce the overhead.

\subsection{Discussion}
\label{subs:dropping}

This section shares a failed trial. 
Inspired by the traditional iterative graph computing algorithms, such as, delta-step Pagerank~\cite{zhang2013maiter} and Single Source Shortest Path~\cite{meyer1998delta,song2018start}, 
we falsely assume the tokens that already converged will no longer change their topics. 
Therefore, our naive design introduces a tracker array to indicate whether the topic of a token remains unchanged for several iterations. If so, this naive dropping method will not sample this token in the following iterations.

\begin{figure}[h]
	\centering
	\vspace{-.1in}
	\includegraphics[width=0.4\textwidth]{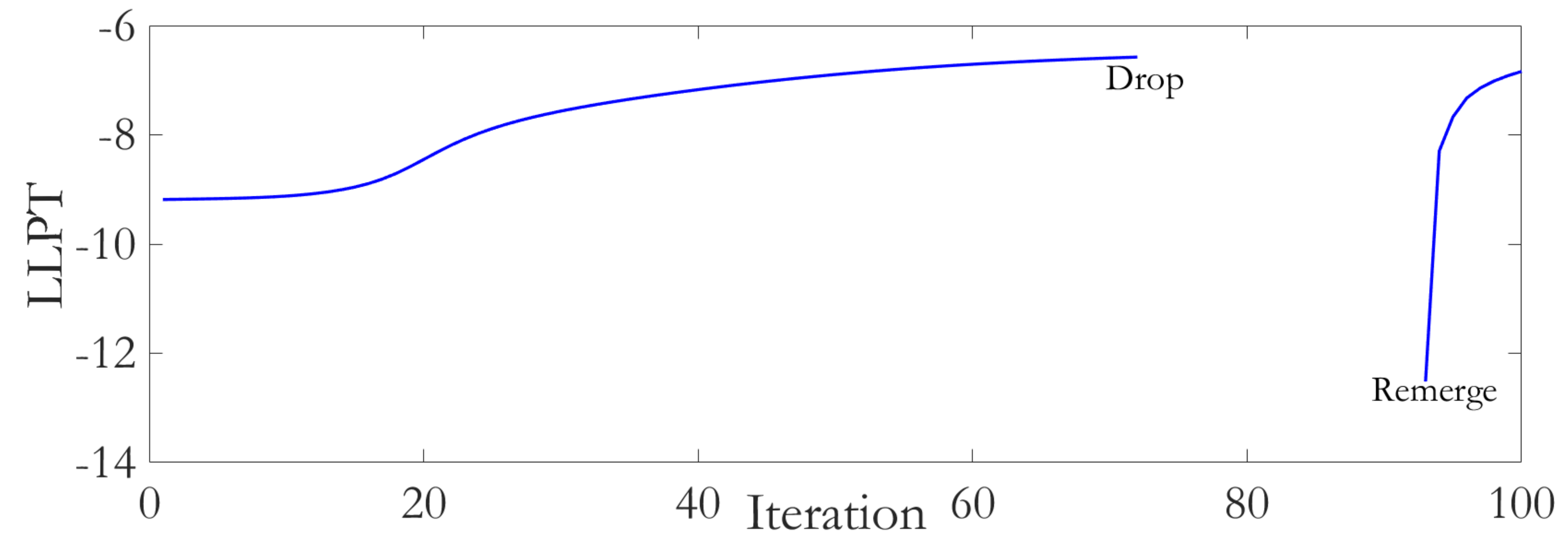}
	\caption{Perplexity of naive dropping strategy on PubMed.
	}
	\label{fig:remerge}
\end{figure}


However, the naive dropping strategy fails to work mainly because it betrays the nature of LDA. Particularly, the core of LDA, i.e., Bayes model, is that the sampling process of LDA is a non-deterministic process. That is, even if the topic of a token remains unchanged for several iterations, which means the probability of assigning this specific token to the same topic is very close to 1, this token still has a chance to change topics because the random number generated from $U[0,1]$ might fall in other topics whose probabilities are low.



Figure \ref{fig:remerge} shows the failure of the naive dropping strategy. 
In this test, the dropping starts at iteration 72. At iteration 90, all dropped tokens are re-included in the training to check whether this strategy works.
Clearly, the results are not good. At the point of re-including, perplexity becomes even smaller than the value before dropping and severely deviates from the correct convergence point. 


\section{Sparsity-Aware Optimization}
\label{sec:sparse}


Reducing the sampling time is important for LDA, so does the space consumption.
This section introduces the sparsity-aware storage format for both \textbf{D} and \textbf{W}, as well as our new mechanisms to facilitate rapid sampling and updating dwelling on these sparse formats.



%

\subsection{Observations}

The space problem faced by {\lda} appears for two types of data, that is, corpus data and algorithmic data. Corpus data is concerned with the gigantic corpus size while algorithmic data is related to both corpus size and number of topics. 
Below, we discuss the details surrounding these two challenges.

The space consumption incurred by the large corpus can be tackled by simply partitioning the corpus into multiple chunks. This way, each GPU will need much less memory for corpus. Doing so, however, comes with one obvious drawback - one needs to repeatedly load each chunk in and out the GPU, which could introduce overhead. {\lda} uses asynchronous kernel launching and memory transfer to hide this cost, similar to existing work~\cite{han2017graphie,holmes2019grnn}. 


In fact, curbing the space consumption of algorithmic data (e.g. \textbf{D} and \textbf{W}) is even more imperative. Below, we unveil the reason from column and row perspective of a matrix.
First, with the climbing of the corpus size, the diversity of the tokens will also increase, indicating the need of a larger number of topics (i.e., number of columns in \textbf{D} and \textbf{W}).
Second, 
for a corpus with abundant documents or unique words, the number of rows in both \textbf{D} and \textbf{W} will also soar.

\begin{figure}[t]
	\centering
	\includegraphics[width=0.4\textwidth]{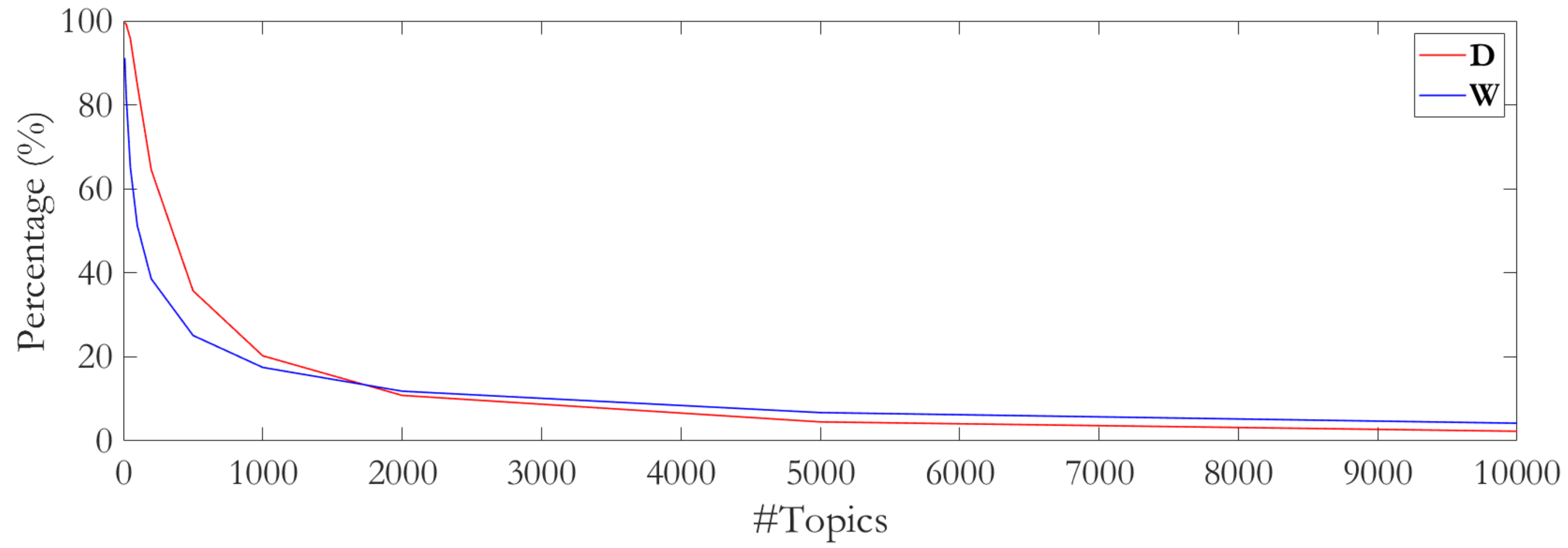}\vspace{-.05in}
	\caption{Density of \textbf{W} and \textbf{D} on NYTimes dataset.\vspace{-.2in}}
	\label{fig:sparsity}
\end{figure}

The good news is that both \textbf{D} and \textbf{W} are often very sparse because very few, if any, of the words or documents will occupy all the topics. 
As shown in Figure~\ref{fig:sparsity}, the {density} of \textbf{D} and \textbf{W} decreases rapidly 
along with the increase of number of topics for the NYTimes dataset. 




\subsection{Sparsity-Aware Representation}
%


We introduce compressed CSR format to store the sparse \textbf{W} and \textbf{D} matrices.
While the traditional CSR contains three major components: \textit{row offset, column indices} and \textit{values}, we further compress \textit{column indices} along with \textit{values} in order to save space and improve performance. 
Knowing that the maximum number of topics are seldom larger than 65,536, 16 bits are enough for storing a single column index or value. Inspired by the pair-storage in Section~\ref{sec:overhead}, {we compress the column indices and values of CSR into and a single integer, where the higher and lower 16 bits are for column index and corresponding value, respectively.}

Storing the entire \textbf{W} in sparse format might not always save space. Particularly, despite sparse format will save memory space for sparse rows in \textbf{W}, words with large number of topics (i.e. dense rows) will, unfortunately, suffer from extra space consumption because CSR requires to store the column indices. 
In contrast, dense format only needs to store the values since the position of the value can automatically indicate its column location. 

We thus advocate to store \textbf{W} in hybrid format. That is, the rows with large volume of nonzero columns (i.e., topics) will be stored in dense format while the remaining rows in sparse format. 
{\lda} comes up with a light-weighted heuristic to estimates the upper bound of \textbf{W} in order to decide whether we store a corresponding row in sparse or dense format. That is, the maximum number of topics one word can possess will not go beyond the number of tokens this word has in the entire corpus. With this rule, one can assign the words with tokens that is larger than the assumed number of topics (i.e. $k$) as dense row and the remaining rows to be the same as the number of tokens. 



To enjoy the space saving from the hybrid format, we propose to group dense words together in token list T. Toward that end, the word identities (IDs) are relabeled based upon their token counts. Basically, words with larger number of tokens hold smaller IDs. Further, words with token count more than the topic number are stored in dense format in \textbf{W}. Subsequently, in each chunk from the token list, we remap the word IDs from the token list into $T_{dense}$ and $T_{sparse}$, respectively, which represent the dense and sparse parts of T, respectively. 
In summary, this hybrid approach comes with the following two advantages. First, comparing to the dense or sparse alone approach, the proposed method will yield the most space saving. Second, storing dense rows into dense format explicitly will reduce the overhead of $\widehat{\textbf{W}}[v]\circ \textbf{D}[d]$.

\subsection{Sparsity-Aware Computation}
\label{SparsityAwareComputation}


Once storing \textbf{W} and \textbf{D} in sparse format is resolved, conducting updating and sampling atop the sparse \textbf{W} and \textbf{D} become a ground challenge for two reasons. First, 
{during sampling, 
we need to do element-wise product of $\widehat{\textbf{W}}[v]\circ \textbf{D}[d]$. 
Given the elements from the same storage index of sparse $\widehat{\textbf{W}}[v]$ and \textbf{D}[d] are most likely not from the same column, we will need to match their columns. }
Second, {to update an element in sparse matrix, we must first find the correct row and column to write the update. So unlike dense matrix, it is impossible to update \textbf{W} once after a new topic is known in sampling kernel. The update of \textbf{W} can only be done by reconstructing from T after sampling, which will consume much more time.}

A naive design could easily combine sampling with updating via keeping a canonical copy of \textbf{W}. During sampling, this design will compute the S' and Q', thus the ratio $t_1=\frac{S'}{S'+Q'+M}$ and  $t_2=\frac{Q'}{S'+Q'+M}$. Based upon the generated random number u, this design could determine to sample from either S' or Q' tree. After arriving at the updated topic for a token, we can immediately update the canonical copy.

However, this naive design also faces two challenges. First, keeping extra canonical copies for \textbf{W} will consume more space. Second, it is hard to predict, for a random token, where to update the canonical copies of \textbf{W} providing they are in the sparse format. Third, given LDA is memory intensive, reading \textbf{W} twice (one for sampling, the other for updating) will hamper the performance.

We only keep a canonical copy for \textbf{W$_{dense}$} and reconstruct \textbf{W$_{sparse}$} as well as the entire \textbf{D} from T after sampling. Below, we discuss the details.
{\lda} keeps a canonical copy of \textbf{W$_{dense}$} for update because the words in the dense rows often contain exceeding number of tokens which span across multiple chunks. In that context, we would need to transfer a large number of chunks if we choose to reconstruct \textbf{W$_{dense}$} from T. In contrast, with a canonical copy of \textbf{W$_{dense}$}, the update to \textbf{W$_{dense}$} can be done quickly because \textbf{W$_{dense}$} is in dense format, as well as in memory during sampling.

\begin{figure}[t]
	\centering
	\includegraphics[width=0.4\textwidth]{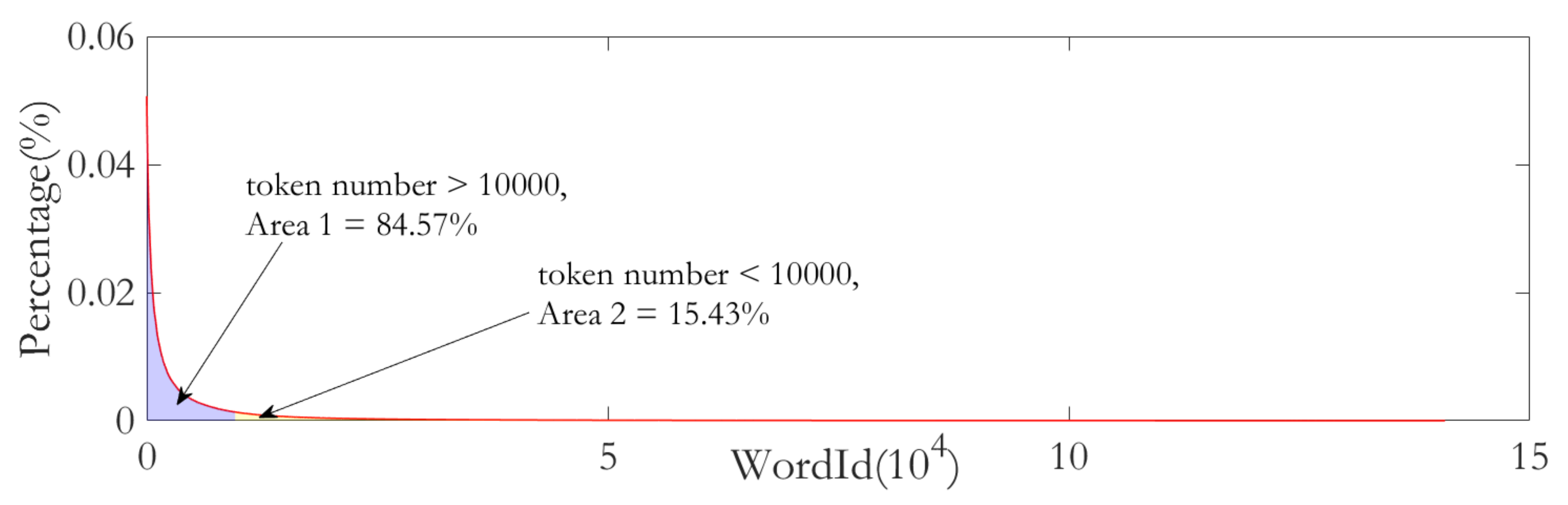}\vspace{-.1in}
	\caption{Token distribution of PubMed dataset. \vspace{-.2in}}
	\label{fig:tokenDistribution}
\end{figure}


For \textbf{W$_{sparse}$}, we only need to read in the \textbf{T$_{sparse}$} part of the token list. Since more than 80\% tokens contribute to \textbf{W$_{dense}$}, \textbf{T$_{sparse}$} will be relatively small. Thus reconstructing \textbf{W$_{sparse}$} will be very fast.

Since a majority of the tokens belong to \textbf{W$_{dense}$} which is updated during sampling, the update of the entire \textbf{W} usually consumes very short time.

The update of \textbf{D} is aided by the inverted index which is discussed in Figure~\ref{fig:update}(b). 
In order to discuss the update to \textbf{D}, we need to understand how {\lda} partitions and preprocesses the corpus. Particularly, each document is solely assigned to one chunk, and all the tokens in each chunk are sorted by wordId.  
Here, assuming this chunk contains three documents of a corpus. This way, we can reconstruct three rows of \textbf{D} with this chunk. Inside of each chunk, the tokens are sorted by wordId for ease of update of \textbf{W$_{sparse}$}. Towards updating \textbf{D}, we resort to the inverted index in Figure~\ref{fig:update}(b). Particularly, one can scan through the CSR to decide which tokens are needed to update rows 0, 1 and 2 of \textbf{D}.

With the updating process being taken care of, sampling becomes the immediate bottleneck. To facilitate a fast S' tree construction, which involves the HP between two CSR rows of \textbf{W} and \textbf{D}, we reconstruct the entire row of \textbf{W} into dense format in shared memory. Afterwards, HP is done by scanning through the specific row of \textbf{D} and use the column index to access that of the dense \textbf{W}. Note, this shared memory will be repeatedly used for all rows of \textbf{W}.

\section{Scalable {\lda} on GPUs}
\label{sec:workload}

This section discusses novel techniques we exploit to better scale {\lda} across GPU threads, as well as GPUs.

\subsection{Intra-GPU Workload Balancing}

For a corpus, the number of tokens per word often follows power law, that is, {a few high frequent words occupy majority of tokens}, as shown in Figure~\ref{fig:tokenDistribution}. The workloads associated with various words are hence largely unbalanced. 
However, the contemporary LDA projects~\cite{SaberLDA} typically assign a block to a word, regardless of the associated workload, leading to severe workload imbalance issue in LDA training. 
This section thus introduces two methods to overcome the workload imbalance problem, that is, \textit{dynamic workload balancing} for small words and \textit{workload splitting} for large words.

\textbf{Small word}. Given various words come with different number of tokens, 
we adopt the dynamic workload balance strategy from a recent work~\cite{gaihre2019xbfs} to address the inter-word workload imbalance issue. 
Note, instead of each block processing the words in a pre-determined manner, this approach will use \textbf{atomicAdd}() to, on-the-fly, determine which word will be processed by the available thread block. 

\begin{figure}[t]
	\centering
	\includegraphics[width=0.36\textwidth]{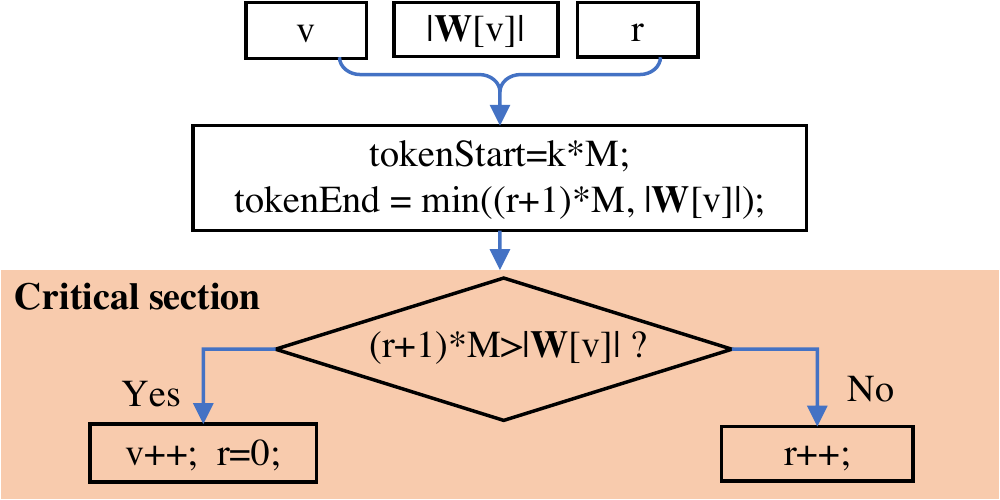}
	\caption{Hierarchical workload balancing. Note, $v$,  $|\textbf{W}[v]|$, $r$ and $M$ are the word index, number of tokens for word $v$, region index, and maximum number of token processed by each index increment, respectively.\vspace{-.15in}}
	\label{fig:hierarchicalWorkload}
\end{figure}

\textbf{Large word}. While applying the dynamic workload balance strategy can largely address the inter-word workload imbalance problem faced by \textit{small words}, it will not work for \textit{large words} which govern too many tokens. In this context, the block that processes the extremely large words will become the straggler. For instance, assuming one corpus has 128 tokens and the most frequent word holds 50 tokens, we use 8 blocks for training. Then, each block should process 16 tokens in workload balanced setting. However, with the dynamic workload balance strategy, the block that processes the word with 50 tokens will be responsible for this entire word alone, leading to workload imbalance.

In order to solve this problem, we introduce \textit{large word dissection}, i.e, very high frequency words are partitioned and processed by multiple thread blocks. Particularly, we can quickly derive the maximum number of tokens a block can process through dividing the total amount of workloads by the number of thread blocks. If the token number of a specific word is larger than this maximum value, we will partition this word into several parts and assign them to multiple blocks. 
{In this work, we use 10,000 as the threshold for {\lda}.}


It is important to note that applying dynamic small word workload balancing and large word dissection together will pose challenges for word assignment. For instance, we need to decide which word and what portion of that word are the next workload. {\lda} introduces a two-level index strategy to deal with this challenge.


Figure \ref{fig:hierarchicalWorkload} shows the design of our two-level index strategy. Word index $v$ determines the word to be processed and region index $r$ determines which region of tokens in that word should be processed. Apparently, the increment of $v$ and $r$ are correlated and must be executed atomically. Considering an atomic function can only be used for a single operation, we propose to use critical section to fulfill that goal. To remedy the absent of critical section support on GPU, {\lda} relies on atomic operations to build a critical section \cite{criticalsection}.

\subsection{Multi-GPU {\lda}}

As the size of corpus and number of topics continue to grow, the training time of LDA also prolongs, which leads to our support of multi-GPU {\lda}. When extending to multiple GPUs, {\lda} is concerned with two essential data structures, that is, data (i.e., T) and algorithmic data (i.e., \textbf{D} and \textbf{W}), and the correlated workload partition, and communication.
 
The good news is that T and \textbf{D} are well partitioned in the single GPU-based design, as discussed in Section~\ref{sec:sparse}. Particularly, each chunk is responsible for similar number of documents. This partition of T leads to evenly partitioned \textbf{D} across GPUs. 
And surprisingly, each chunk actually contains a similar number of tokens. Using UMBC dataset on four GPUs as an example, the maximum and minimum workload chunks only have a difference of 5\% in terms of the number of tokens.


For word topic matrix, i.e., \textbf{W}, unlike the single GPU version, we keep an in-memory canonical copy for both $\textbf{W}_{dense}$ and $\textbf{W}_{sparse}$ given we have adequate space for all the data structures. After all chunks are processed, we can update both $\textbf{W}_{dense}$ and $\textbf{W}_{sparse}$ by summing up the canonical copies across all GPUs and broadcasting the result back to all of them.

\section{Experiments}
\label{sec:eval}

We implement {\lda} with $\sim$4,000 lines of C++/CUDA code and compile the source code with Nvidia CUDA 9.2 toolkit and -O3 optimization compilation flag. 
We use two platforms to evaluate {\lda}. {For comparison with state-of-the-art SaberLDA, we use an Nvidia GTX 1080 GPU - identical platform used in SaberLDA - on an Alienware with 24 GB memory and Intel(R) Core(TM) i7-8700 (3.20Hz) CPU.
For {\lda} internal study, we use a customized server which installs a dual-socket Xeon processor with 24 cores, and four Nvidia V100 GPUs}. {Note, each reported result is an average of five runs, where the differences across various executions are very small ($<$ 1\%).}

\textbf{Dataset}. We evaluate {\lda} with two popular datasets that are also studied by cuLDA~\cite{culdacgs} and SaberLDA~\cite{SaberLDA}:
\begin{itemize}
    \item PubMed~\cite{Dua:2019}: 8,200,000 documents, 141,043 unique words and 738M tokens.
    \item NYTimes~\cite{Dua:2019}: 299,752 documents, 101,636 unique words and ~100M tokens. 
\end{itemize}

To better study the scalability and real-world impacts, we further prepare the following dataset 
by {text splitting, stop words removing and non-frequent words stemming}:
\begin{itemize}
    \item UMBC: 40,000,000 documents, 200,000 unique words and 1.33 billion tokens. This dataset is obtained from UMBC webbase corpus~\cite{UMBC}.
\end{itemize}

\vspace{-.1in}



\subsection{{\lda} vs. State-of-the-art}

\definecolor{Gray}{gray}{0.9}

\begin{table}[t]
\centering  
{\small
	\begin{center}
    \setlength{\tabcolsep}{1.5 mm}
  \scalebox{0.9}{
	\begin{tabular}{r|r|r|r|r|r}
\hline
\multicolumn{1}{c|}{Method}& \multicolumn{1}{c|}{\#Topics} &\multicolumn{1}{c|}{\textbf{W}}  &\multicolumn{1}{c|}{\textbf{D}} & \multicolumn{1}{c|}{T}      & \multicolumn{1}{c}{Total}   \\ \hline
\multirow{3}{*}{{cuLDA/SaberLDA}} & 1,000    & 1.08 GB     & 1.45 GB     & 2.16 GB & 4.69 GB \\ \cline{2-6} 
                                                 				&10,000    &10.8 GB    &1.45 GB     &2.16 GB  &14.41 GB \\ \cline{2-6} 
                                                  				&32,768    &35.4 GB    &1.45 GB     &2.16 GB  &39.01 GB \\ \hline
\multirow{3}{*}{\lda}             			  & 1,000      & 0.31 GB     & 0.98 GB     & 4.97 GB   & 6.26 GB  \\ \cline{2-6} 
                                                   				& 10,000    & 1.63 GB     & 0.98 GB     & 4.97 GB   & 7.58 GB  \\ \cline{2-6} 
                                                   				&32,768    &2.5 GB    &0.98 GB     &4.97 GB  &8.44 GB \\ \hline
\end{tabular}
}
	\end{center}
	}
		\caption{{\lda} vs {cuLDA/SaberLDA} memory consumption on PubMed dataset. The corpus is partitioned into 8 chunks during computation.
		\vspace{-.05in}}
	\label{table:mem}
\end{table}





\begin{figure}[htbp]
	\centering
	\vspace{-.1in}
	\includegraphics[width=0.45\textwidth]{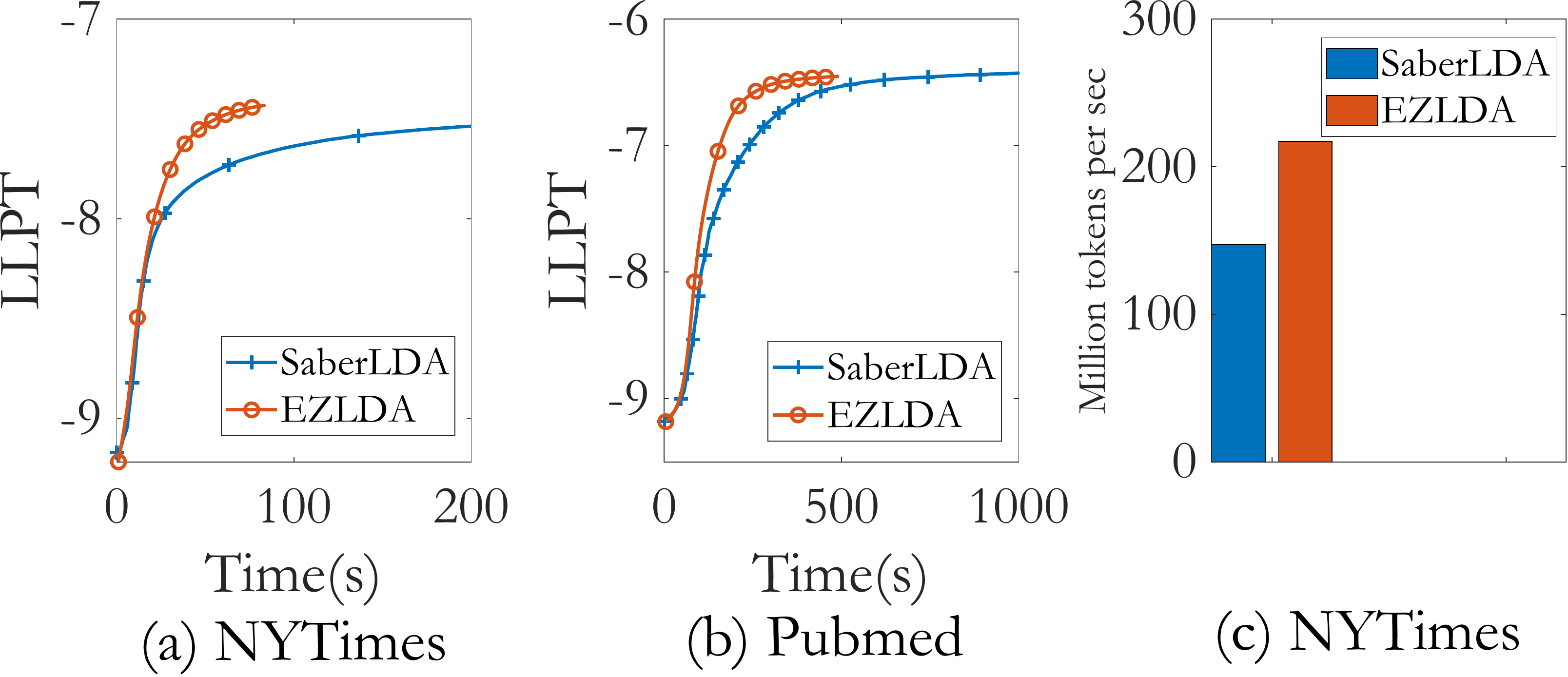}
	\caption{(a) The convergence of {\lda} vs SaberLDA with 1,000 topics on NYTimes dataset on Titan 1080. (b) The convergence of {\lda} vs SaberLDA with 1,000 topics on Pubmed dataset  on Titan 1080. {(c) Throughput of {\lda} vs SaberLDA for first 100 iterations on NYTimes on Titan 1080.}
	\vspace{-.2in}
	}
	\label{fig:SaberLDAvsScaLDA}
\end{figure}

\begin{figure}[htbp]
	\centering
	\includegraphics[width=0.4\textwidth]{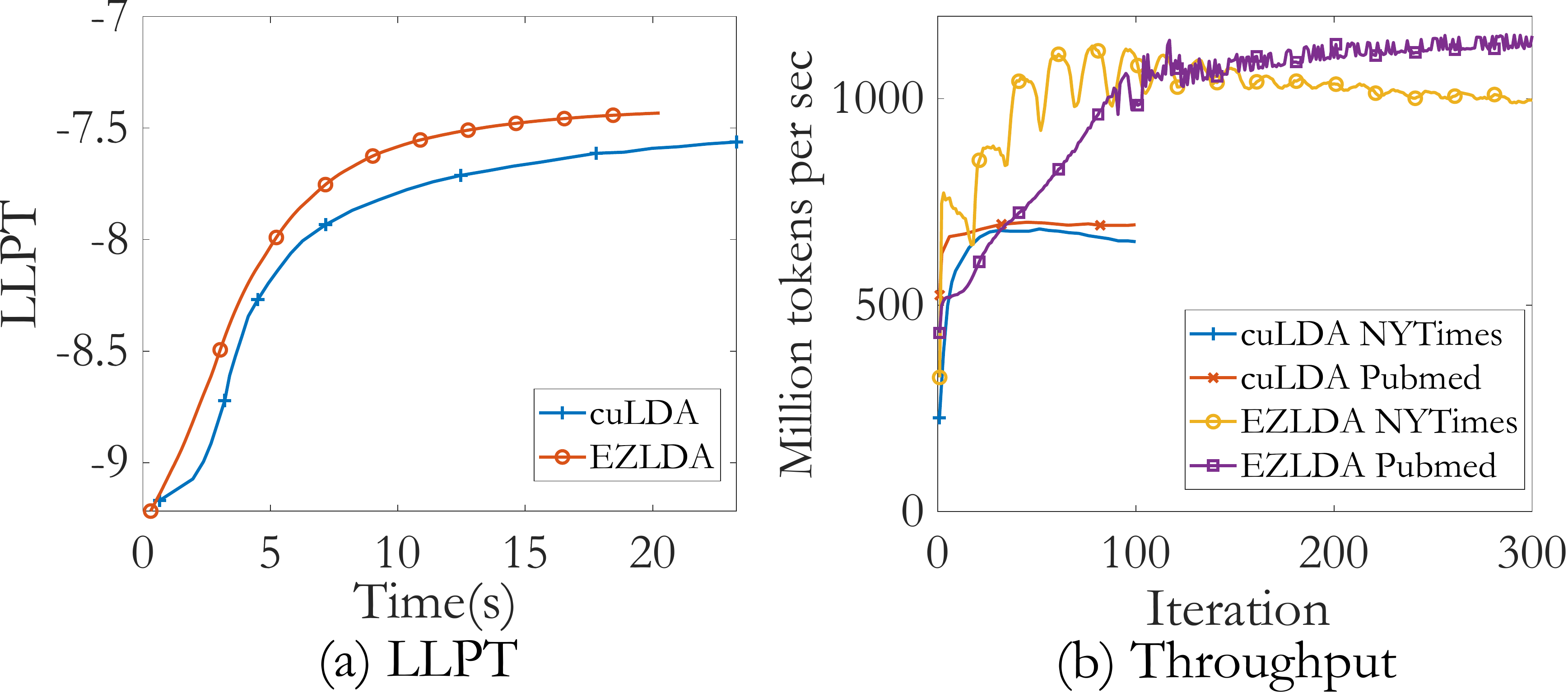}
	\caption{{\#Tokens/second for {\lda} vs. cuLDA on V100:(a) NYTimes, (b) Pubmed, (c)LLPT on Nytimes}
	\vspace{-.2in}
	}
	\label{fig:cuLDAvsScaLDA}
\end{figure}

{Table~{\ref{table:mem}}, Figures~{\ref{fig:SaberLDAvsScaLDA}} and~{\ref{fig:cuLDAvsScaLDA}} compare {\lda} against the state-of-the-art, i.e., SaberLDA and cuLDA for space complexity, convergence speed and throughput (\#tokens/second), respectively.} 

\textbf{Space.}
As shown in Table~\ref{table:mem}, {\lda} consumes 33\% more space for small \#topics = 1,000 compared with SaberLDA and cuLDA. But we save 47\% and 78\% space when \#topics is large, e.g., 10,000 and 32,768. Particularly, {\lda} require more space than SaberLDA {and cuLDA} for T because we need to allocate space in T for K$_1$/K$_2$ pair (Section~\ref{sec:overhead}) and $M$ (Equation~\ref{con:MValue}). However, {\lda} manages to save much more memory on \textbf{W} thanks to sparsity aware representation. 

{\textbf{{\lda} vs SaberLDA.}  
Figure~{\ref{fig:SaberLDAvsScaLDA}} compares {\lda} against SaberLDA on convergence speed and throughput, i.e, \#tokens/second. Since SaberLDA is not open source, we cite the performance numbers from their manuscript and run {\lda} on identical GPU for fair comparison. As shown in Figures~{\ref{fig:SaberLDAvsScaLDA}}(a) and~{\ref{fig:SaberLDAvsScaLDA}}(b), {\lda} climbs to higher perplexity with shorter training time. 
For throughput, as shown in Figure~{\ref{fig:SaberLDAvsScaLDA}}(c), {\lda} achieves 1.5$\times$ speedup, on average, for the first 100 iterations on NYTimes. Note, since SaberLDA does not include the \#tokens/second statistics, we follow cuLDA to derive this number for NYTimes according to Figure 9 in SaberLDA~{\cite{SaberLDA}}.}

{\textbf{{\lda} vs cuLDA.}  
Though cuLDA outperforms SaberLDA, {\lda}, as shown in Figure~{\ref{fig:cuLDAvsScaLDA}}(a),  still manages to convergence faster than cuLDA on NYTimes (cuLDA does not include LLPT for PubMed). 
Thanks to cuLDA which includes \#tokens/second for both datasets, we report the comparison of this metric in Figure~{\ref{fig:cuLDAvsScaLDA}}(b). Particularly, {\lda} achieves an average throughput of 905 and 770 million tokens/second for the first 100 iterations and retains over 1000 million tokens/sec after 100 iterations on NYTimes and PubMed, respectively. This outperforms cuLDA that retains 633 and 686 million tokens/second, on average, for the first 100 iterations on NYTimes and PubMed, respectively.}

\subsection{{\lda} Performance Study}

\begin{figure}[htbp]
	\centering
	\includegraphics[width=0.4\textwidth]{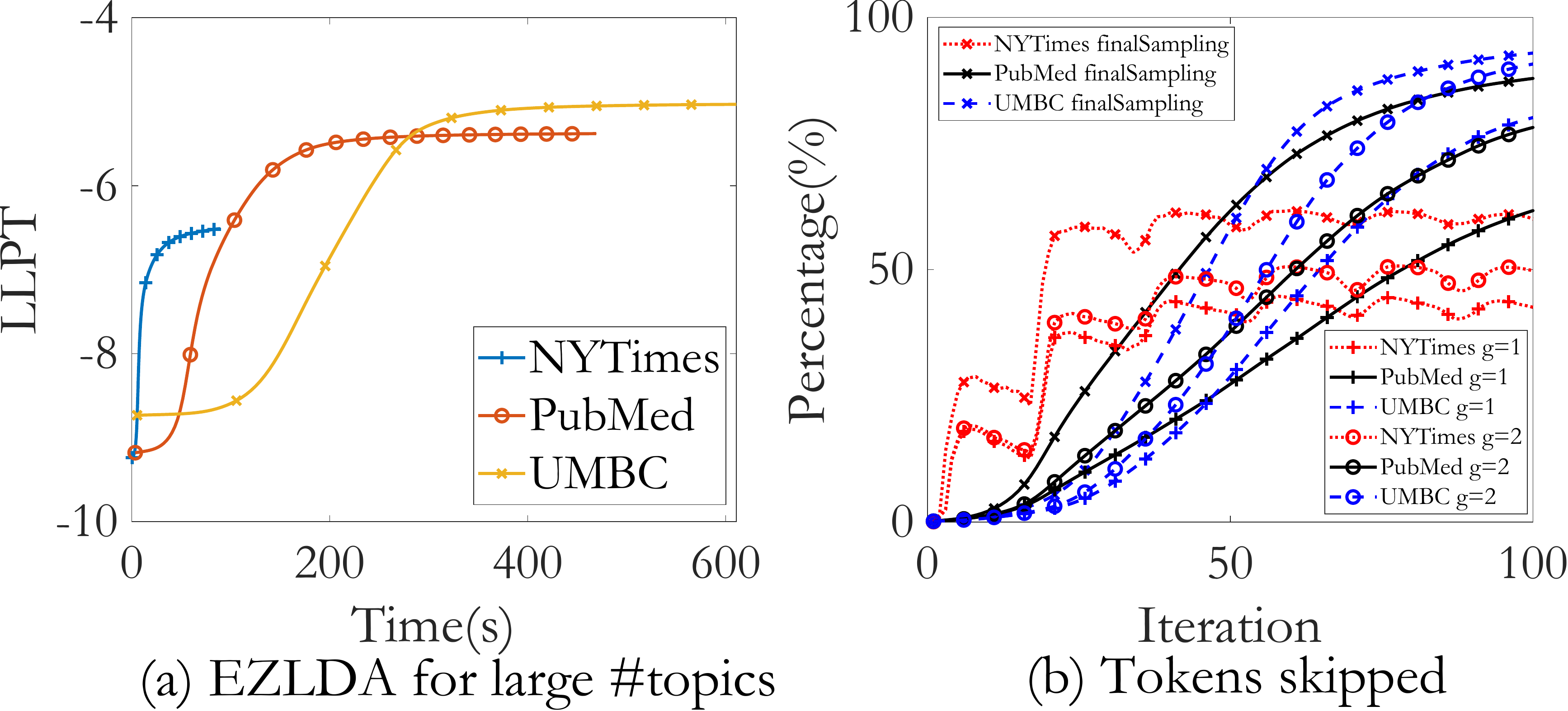}
	\caption{(a) {\lda} for large \#topics, i.e., 32,768. (b) The percentage of tokens skipped by three-branch sampling for \#topics = 1,000, where $g$ is the parameter from Equation~\ref{con:Sest}.
\vspace{-.1in}}
	\label{fig:LargeKThreeBranchCompare}
\end{figure}

\textbf{Large number of topics}. As shown in Figure~\ref{fig:LargeKThreeBranchCompare}(a), {\lda} can handle all the datasets with 32,768 topics on a single GPU. This is an important capability for real-world scale corpus~\cite{LDAStar}. 


\textbf{Three-branch sampling}.
Figure \ref{fig:LargeKThreeBranchCompare}(b) profiles the impact of three-branch sampling. In general, we find this method is more effective when dealing with larger dataset. Particularly, 
{{NYTimes enjoys skipping 60\% of the tokens during the final sampling and nearly 50\% tokens} skip the S construction at iteration 100. For PubMed, 87\% tokens skip the final sampling and nearly 74\% tokens skip the S construction at iteration 100. For UMBC, 93\% tokens skip final sampling and nearly 89\% tokens skip the S' construction at iteration 100.} 
We also study different $g$ in Equation~\ref{con:Sest}. As expected, more tokens are skipping S' construction for larger $g$, because larger $g$ makes $S_{est}$ closer to S'.



\begin{figure}[htbp]
	\centering
	\includegraphics[width=0.4\textwidth]{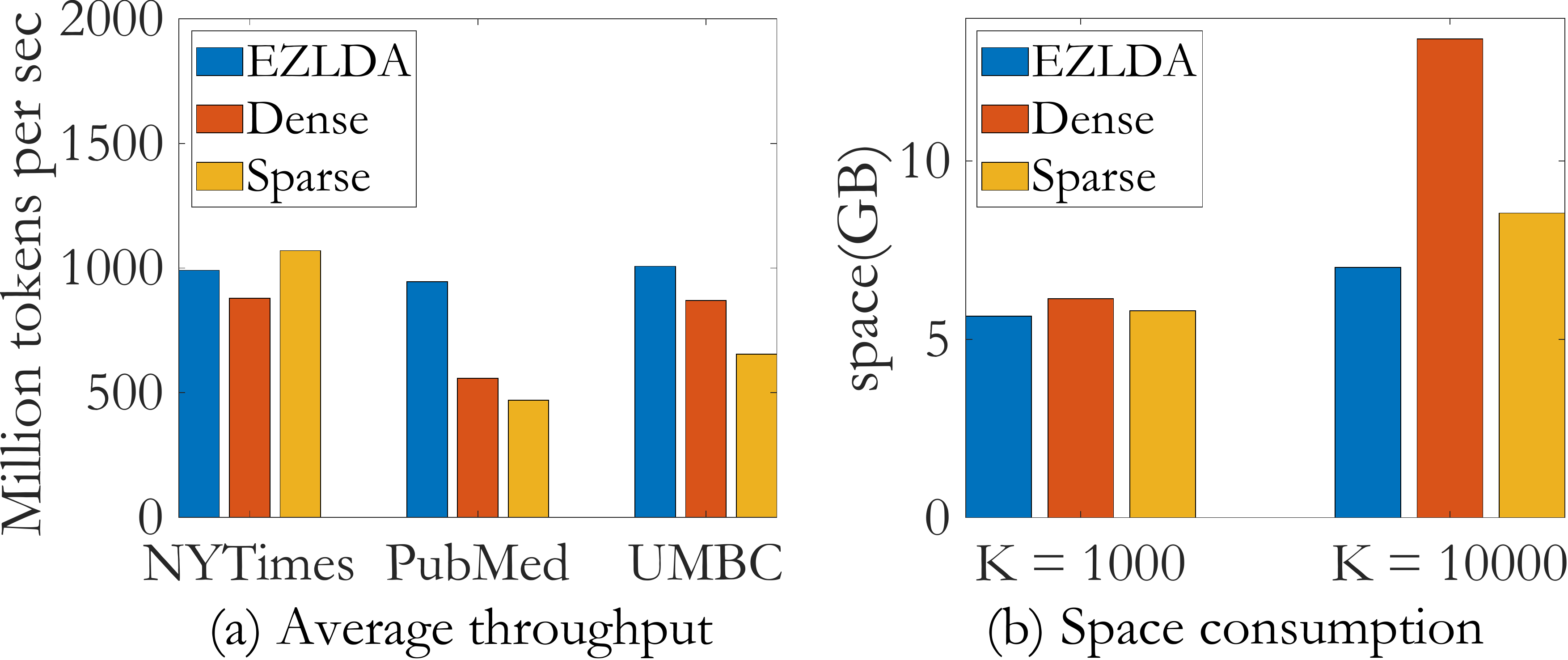}
	
	\caption{{{\lda} hybrid representation vs. dense-only and sparse-only representations for topic = 1,000: (a) throughput and (b) space complexity for UMBC dataset.}
	\vspace{-.1in}}
	\label{fig:denseSparseCompare}
\end{figure}

\textbf{Hybrid storage of W}. Figure~\ref{fig:denseSparseCompare} profiles the {impact of dense/sparse hybrid representations}. 
{The key conclusion is that our hyrbid optimization can both improve performance and save space (at least for the large \#topics case).}
As shown in Figure~\ref{fig:denseSparseCompare}(a), on average, hybrid format is 1.34$\times$ and 1.47$\times$ faster than the dense and sparse only formats, respectively. {Compared with {\lda}, sparse format needs to update \textbf{W} after all chunks are processed, which means all chunks need to be transferred back to GPU a second time to finish the update. For dense format, much time will be wasted on updating rows of \textbf{W} corresponding to small words.} Further, the hybrid format consumes 17.8\% and 47.8\% less space than sparse format and dense format for K = 10,000, respectively.

\begin{figure}[htbp]
	\centering
	\vspace{-.1in}
	\includegraphics[width=0.4\textwidth]{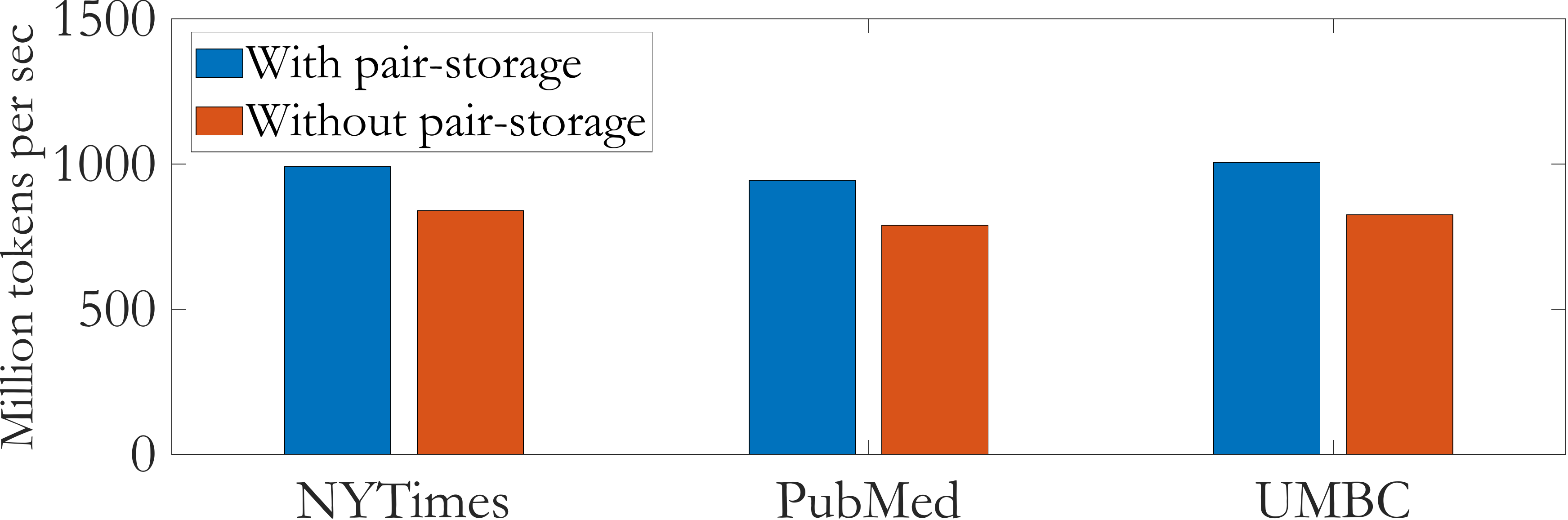}\vspace{-.1in}
	\caption{The performance impacts of pairStorage on V100 GPU for \#topics = 1000.
	\vspace{-.1in}}
	\label{fig:pairStorage}
\end{figure}

\textbf{Pair K1/K2, C1/C2 {and D} storage}, as shown in Figure~\ref{fig:pairStorage}, yields 1.12$\times$, 1.19$\times$ and 1.22$\times$ speedup on NYTimes, PubMed and UMBC datasets, respectively. The speedup is achieved because LDA training is memory-bound~\cite{memoryBound} and pair-storage significantly reduces the global memory traffic in three-branch sampling. 

\subsection{Scalable {\lda}} 


\begin{figure}[htbp]
	\centering
	\includegraphics[width=0.4\textwidth]{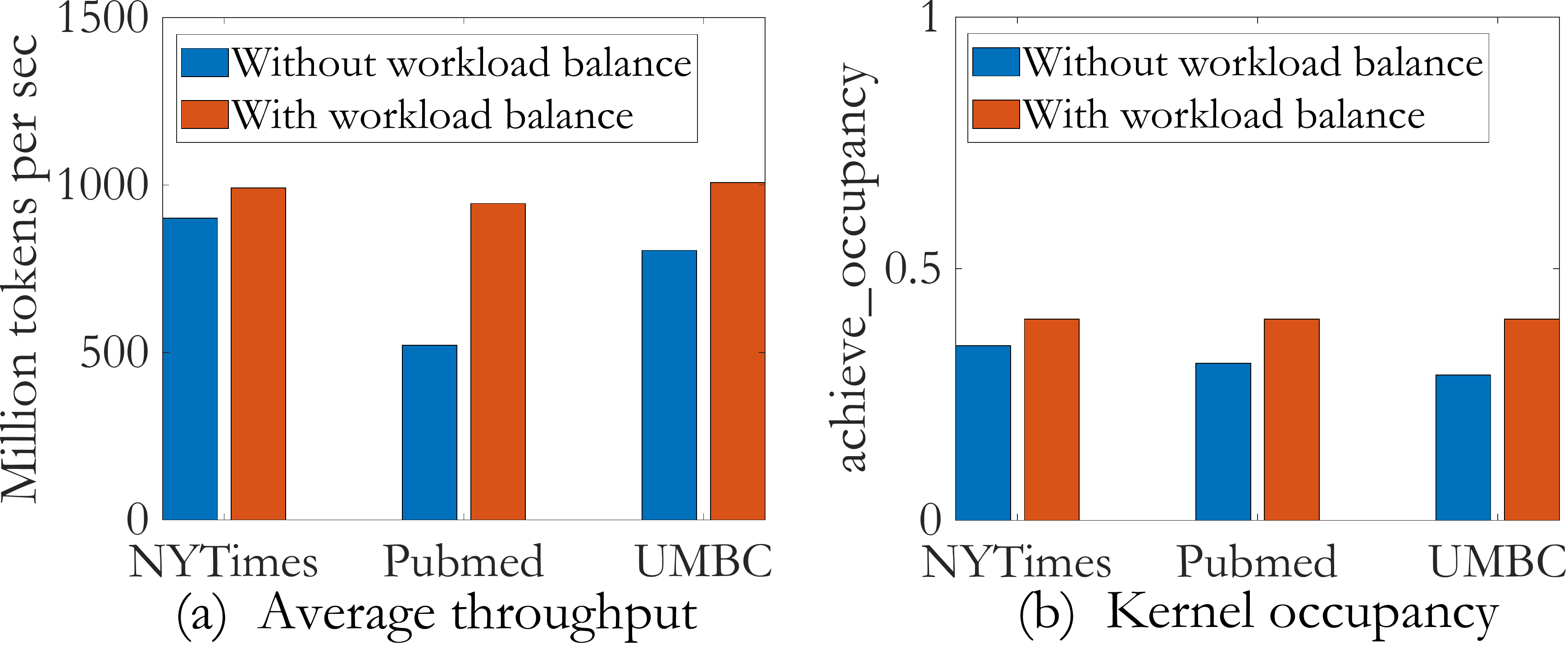}
	\caption{{Profiling the impacts of intra-GPU workload balancing: (a) \#Tokens/second for first 300 iteration (b) achieve\_occupancy.}
    \vspace{-.2in}
	}
	\label{fig:workloadBalance}
\end{figure}

{\textbf{Hierarchical workload balancing}. Using three-branch sampling without workload balance as baseline, as shown in Figure~{\ref{fig:workloadBalance}}(a), on average, our hierarchical workload balancing technique yields 1.1$\times$, 1.7$\times$ and 1.2$\times$ speedup on NYTimes, PubMed and UMBC, respectively, for \#tokens/second. The speedup on PubMed dataset is higher because this dataset presents higher workload imbalance. The speedup is resulted from that workload balancing can improve the GPU occupancy~{\cite{nvidia_nvprof}}. As shown in Figure~{\ref{fig:workloadBalance}}(b), we improve the achieved\_occupancy ratio by 27\% across the datasets. Note, achieved\_occupancy means the ratio of active warps over maximum number of supported warps on the multiprocessor. 
}

\begin{figure}[htbp]
	\centering
	\vspace{-.1in}
	\includegraphics[width=0.4\textwidth]{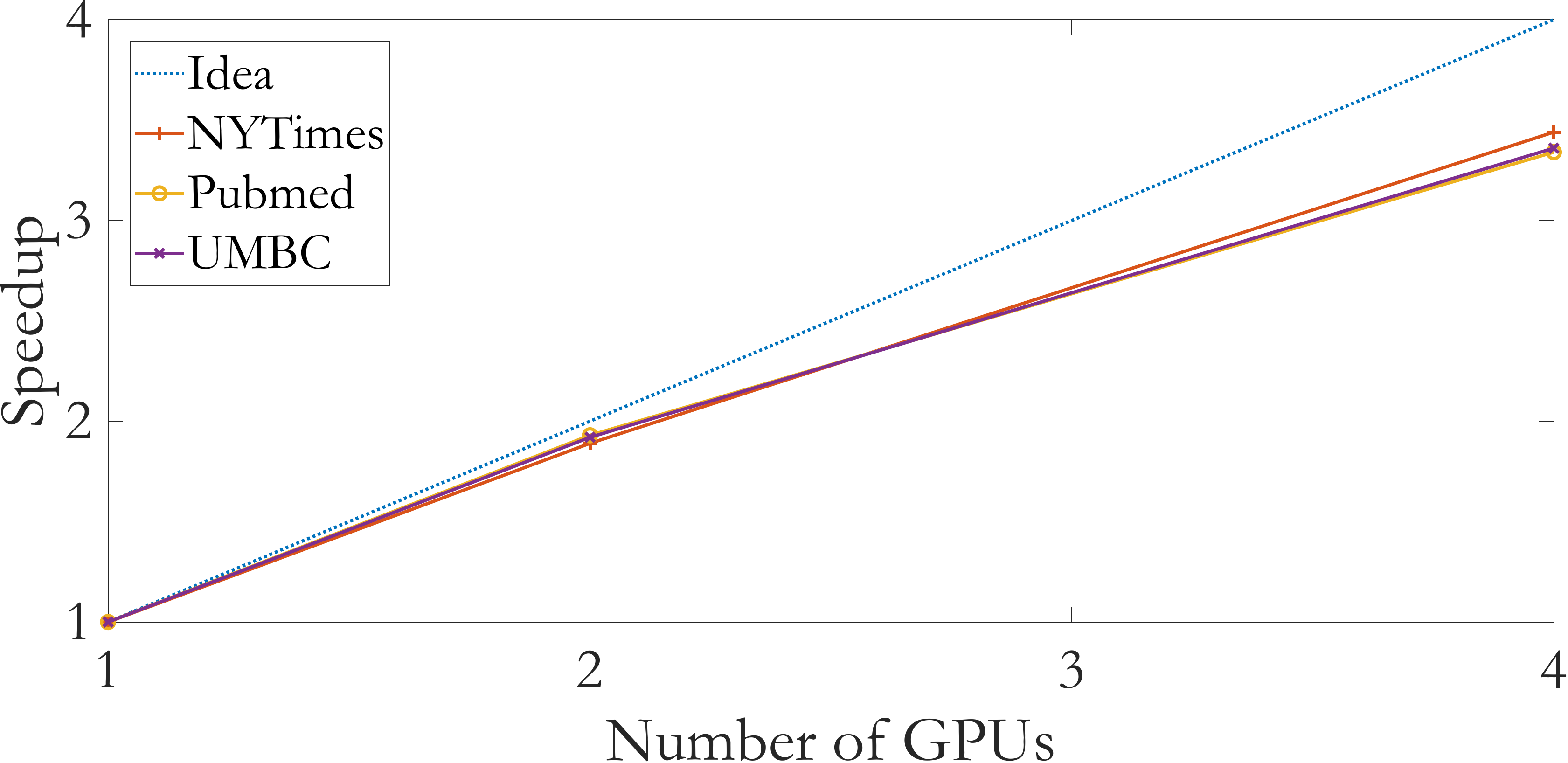}
	\caption{The performance impacts of scalable {\lda}.
    \vspace{-.1in}
	}
	\label{fig:multiGPU}
\end{figure}

\textbf{Multi-GPU scalability}. Figure \ref{fig:multiGPU} shows that {\lda} can scale to four V100 GPUs with {3.44$\times$, 3.34$\times$ and 3.36$\times$} speedup on NYTimes, PubMed and UMBC dataset, respectively. {While this result indicates that {\lda} is scalable, we also notice that {\lda} cannot achieve linear scalability. The reason lies in the need of communicating \textbf{W} and the slight workload imbalance across partitions}.  

\subsection{{Performance Counter and GPU Generation Impacts}}

\begin{figure}[htbp]
	\centering
	\includegraphics[width=0.4\textwidth]{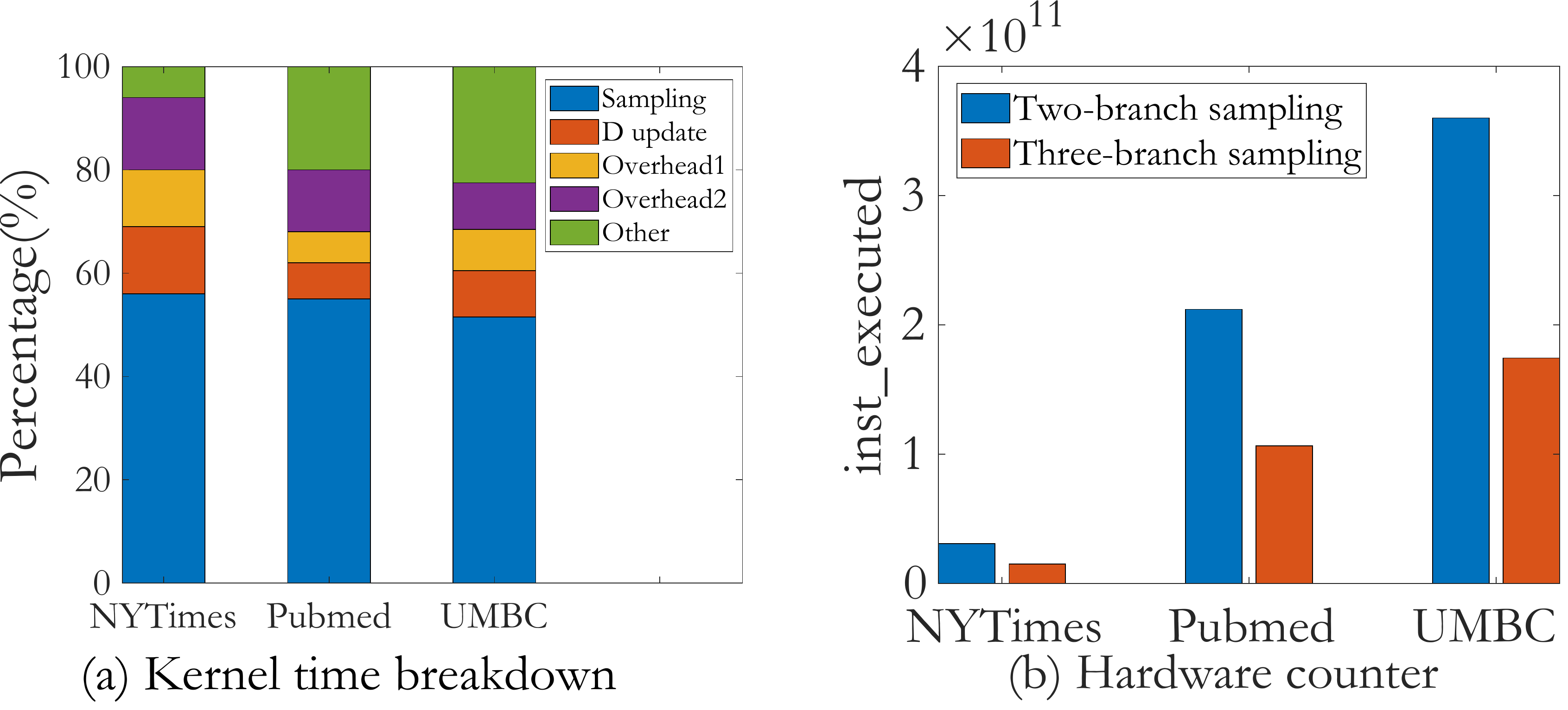}
	\caption{{Profiling {\lda}: (a) kernel time breakdown (b) Hardware counter.} 
	\vspace{-.2in}
	}
	\label{fig:twoThreeBranchCompare}
\end{figure}

{
Figure~{\ref{fig:twoThreeBranchCompare}}(a) studies the time consumption breakdown in three-branch optimization. Though three-branch can skip tremendous tokens, it also introduces two noticeable overheads, i.e., steps} \ballnumber{2} and \ballnumber{3} {in Figure~{\ref{fig:newMethod}}(c). On average, these two steps consume 8\% and 12\% of the total runtime, respectively. 
Figure~{\ref{fig:twoThreeBranchCompare}}(b) further profiles the microarchitectural impacts of three-branch optimization. Particularly, we profile the inst\_executed~{\cite{nvidia_nvprof}} and find that three-branch optimization can reduce the executed instructions by 49\%, on average, across the three datasets.}

\begin{figure}[htbp]
	\centering
	\includegraphics[width=0.45\textwidth]{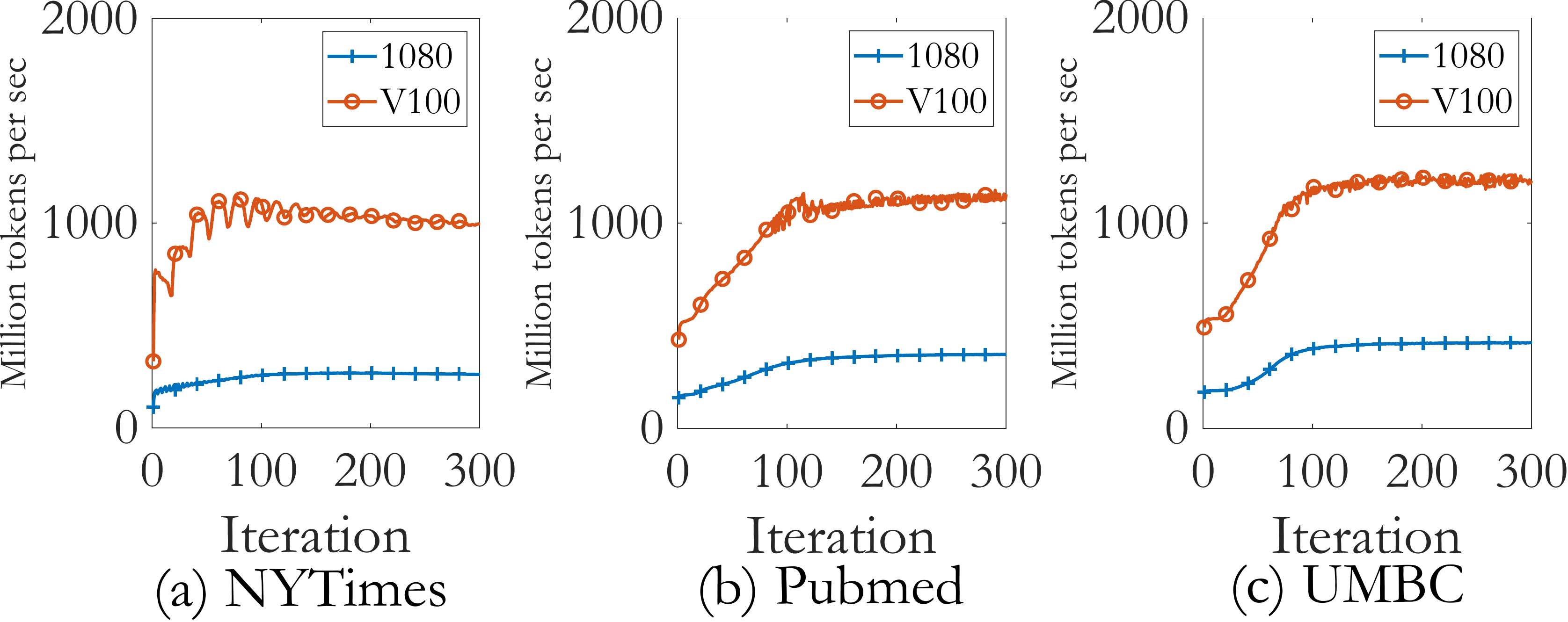}
	\caption{{Throughput impacts of Titan 1080 vs. Volta V100. (a) NYTimes (b) Pubmed (c) UMBC.}
	\vspace{-.2in}
	}
	\label{fig:1080vsV100TokenPerSec}
\end{figure}

\begin{figure}[htbp]
	\centering
	\includegraphics[width=0.45\textwidth]{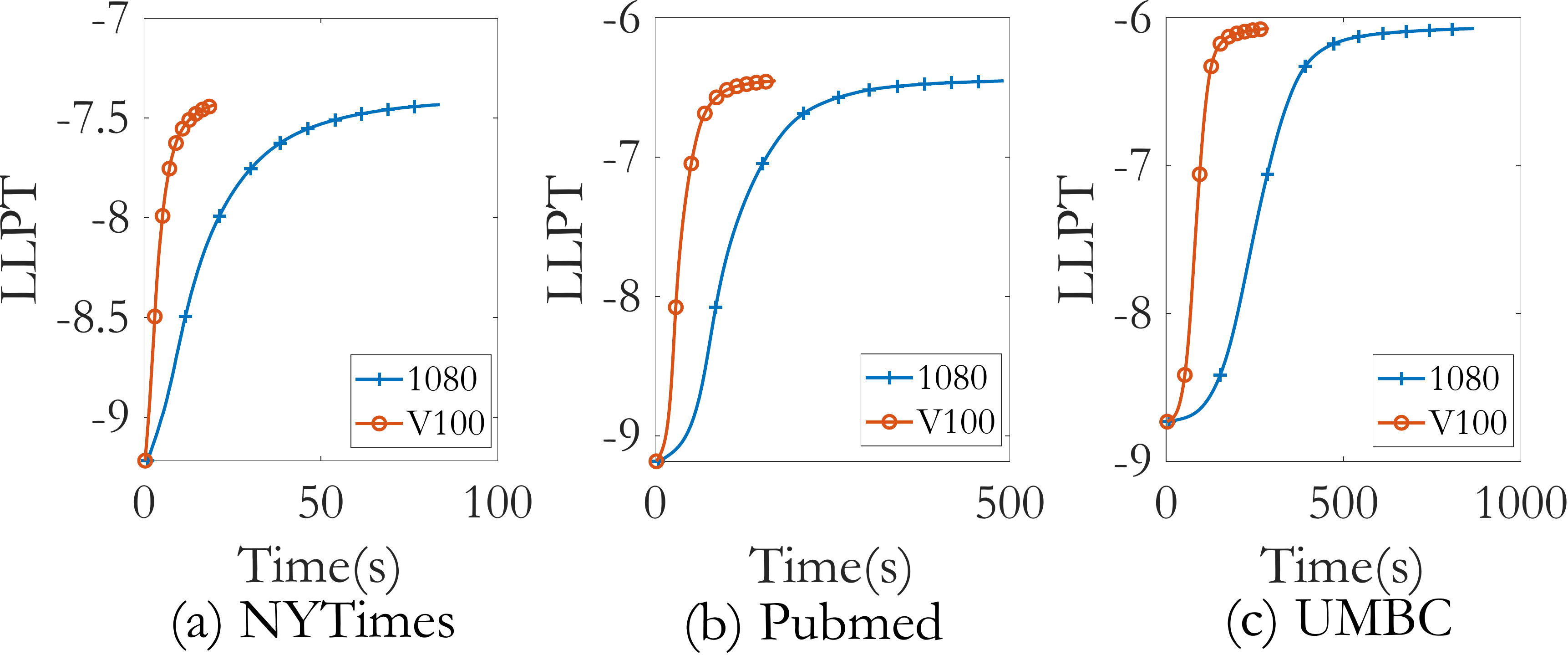}
	\caption{{Convergence speed impacts of Titan 1080 vs. Volta V100. (a) NYTimes (b) Pubmed (c) UMBC.}
	\vspace{-.2in}
	}
	\label{fig:1080vsV100LLPT}
\end{figure}

{Figures~{\ref{fig:1080vsV100TokenPerSec}} and~{\ref{fig:1080vsV100LLPT}} study the GPU generation impacts on \#tokens/second and convergence speed for {\lda}, respectively. Since the bandwidths of Titan 1080~{\cite{nvidia_1080}} and V100~{\cite{nvidia_V100}} are 320 GB/s and 900 GB/s, respectively, we expect the performance impacts would also be around 3$\times$. As shown in Figures~{\ref{fig:1080vsV100TokenPerSec}},
{\lda} can achieve an average of 991, 945 and 1007 million tokens/sec on V100 GPU and 250, 311 and 363 million tokens/sec on Titan 1080 GPU on NYTimes, Pubmed and UMBC datasets respectively. As shown in Figure~{\ref{fig:1080vsV100LLPT}}, {\lda} also converges significantly faster on V100 than Titan 1080.}

\section{Conclusion}
\label{sec:conclusion}

In this paper, we present {\lda} that achieves superior performance over the state-of-the-art attempts with lower memory consumption. 
Particularly, {\lda} introduces a novel three-branch sampling method which takes advantage of the convergence heterogeneity of various tokens to reduce the redundant sampling task. 
Further, to enable sparsity-aware format for both \textbf{D} and \textbf{W} on GPUs with fast sampling and updating, we introduce hybrid format for \textbf{W} along with corresponding token partition to T and inverted index designs. 
Last but not the least, we design strategies to balance workload across GPU threads and scale {\lda} across multiple GPUs. 

\section*{Acknowledgment}

Shilong Wang and Hang Liu contributed equally to this work. Hengyong Yu is the corresponding author.

\bibliographystyle{unsrt}
\bibliography{reference.bib}

\end{document}